\documentclass[aps,pre,onecolumn,superscriptaddress]{revtex4-1}

\pdfoutput=1

\usepackage{amsmath,amsfonts,amssymb,amsthm,color}
\usepackage{graphicx}
\usepackage{hyperref}
\usepackage{ulem} 
\allowdisplaybreaks

\begin{document}

\title{Particle size selection in capillary instability \\ of locally heated co-axial fiber}

\author{Saviz Mowlavi}
\thanks{These authors contributed equally to this work.}
\affiliation{Department of Mechanical Engineering, Massachusetts Institute of Technology, Cambridge, MA 02139, USA}
\affiliation{Laboratory of Fluid Mechanics and Instabilities, Ecole Polytechnique F\'ed\'erale de Lausanne, 1015 Lausanne, Switzerland}
\author{Isha Shukla}
\thanks{These authors contributed equally to this work.}
\affiliation{Laboratory of Fluid Mechanics and Instabilities, Ecole Polytechnique F\'ed\'erale de Lausanne, 1015 Lausanne, Switzerland}
\author{Pierre-Thomas Brun}
\affiliation{Department of Chemical and Biological Engineering, Princeton University, Princeton, NJ 08540, USA}
\author{Fran\c cois Gallaire}
\email{francois.gallaire@epfl.ch}
\affiliation{Laboratory of Fluid Mechanics and Instabilities, Ecole Polytechnique F\'ed\'erale de Lausanne, 1015 Lausanne, Switzerland}

\date{\today}

\begin{abstract}
Harnessing fluidic instabilities to produce structures with robust and regular properties has recently emerged as a new fabrication paradigm. This is exemplified in the work of Gumennik et al.~[Nat.~Comm.~4:2216, DOI: 10.1038/ncomms3216, (2013)], in which the authors fabricate silicon spheres by feeding a silicon-in-silica co-axial fiber into a flame. Following the localized melting of the silicon, a capillary instability of the silicon-silica interface induces the formation of uniform silicon spheres. 
Here, we try to unravel the physical mechanisms at play in selecting the size of these particles, which was notably observed by Gumennik et al.~to vary monotonically with the speed at which the fiber is fed into the flame. Using a simplified model derived from standard long-wavelength approximations, we show that linear stability analysis strikingly fails at predicting the selected particle size. Nonetheless, nonlinear simulations of the simplified model do recover the particle size observed in experiments, without any adjustable parameters. This shows that the formation of the silicon spheres in this system is an intrinsically nonlinear process that has little in common with the loss of stability of the underlying base flow solution.
\end{abstract}

\maketitle

\section{Introduction}

Mechanical instabilities in engineered structures have historically been perceived as failure mechanisms. As such, an enduring motivation for their study has been the desire to avoid them. Recently, however, we have started to witness a paradigm shift wherein structural instabilities are instead sought after due to their natural ability to produce regular patterns that would be difficult or costly to achieve otherwise \citep{reis2015}. Interestingly, this philosophy has been applied for a long time in fluid mechanics, in particular in the field of inkjet printing. First introduced commercially by Siemens in 1951, continuous inkjet printers have long relied on the Rayleigh-Plateau instability \citep{plateau1873,rayleigh1878} to break a liquid jet emerging from a high-pressure reservoir into a multitude of uniformly-sized droplets, some of which are subsequently deflected towards the substrate by means of an electrostatic field \citep{martin2008}.

Returning to solid structures, recent utilization of the solid-liquid phase transition inherent to a wide range of materials has opened new doors by enabling the harnessing of fluidic instabilities, such as the aforementioned Rayleigh-Plateau instability, in order to produce solid structures with robust and regular properties \citep{gallaire2017}. In a seminal contribution, Kaufman \textit{et al.}~\cite{kaufman2012} first adopted this idea and devised a scalable and efficient instability-mediated fabrication process for millimeter to nanometer-sized spherical particles \citep{rotello2004}. The procedure begins with thermal drawing of a co-axial rod into a long and thin fiber consisting of a solid core encased in a cladding of a different material. The fiber is then exposed to a uniform heat source, inducing melting of the core and softening of the outer cladding. This, in turn, triggers a Rayleigh-Plateau instability of the core-cladding interface, which results in global break-up of the continuous core into a regular string of spherical particles. These particles are finally solidified upon cooling of the fiber and released by dissolving the cladding.

\begin{figure}
\centering
\includegraphics[width=\textwidth]{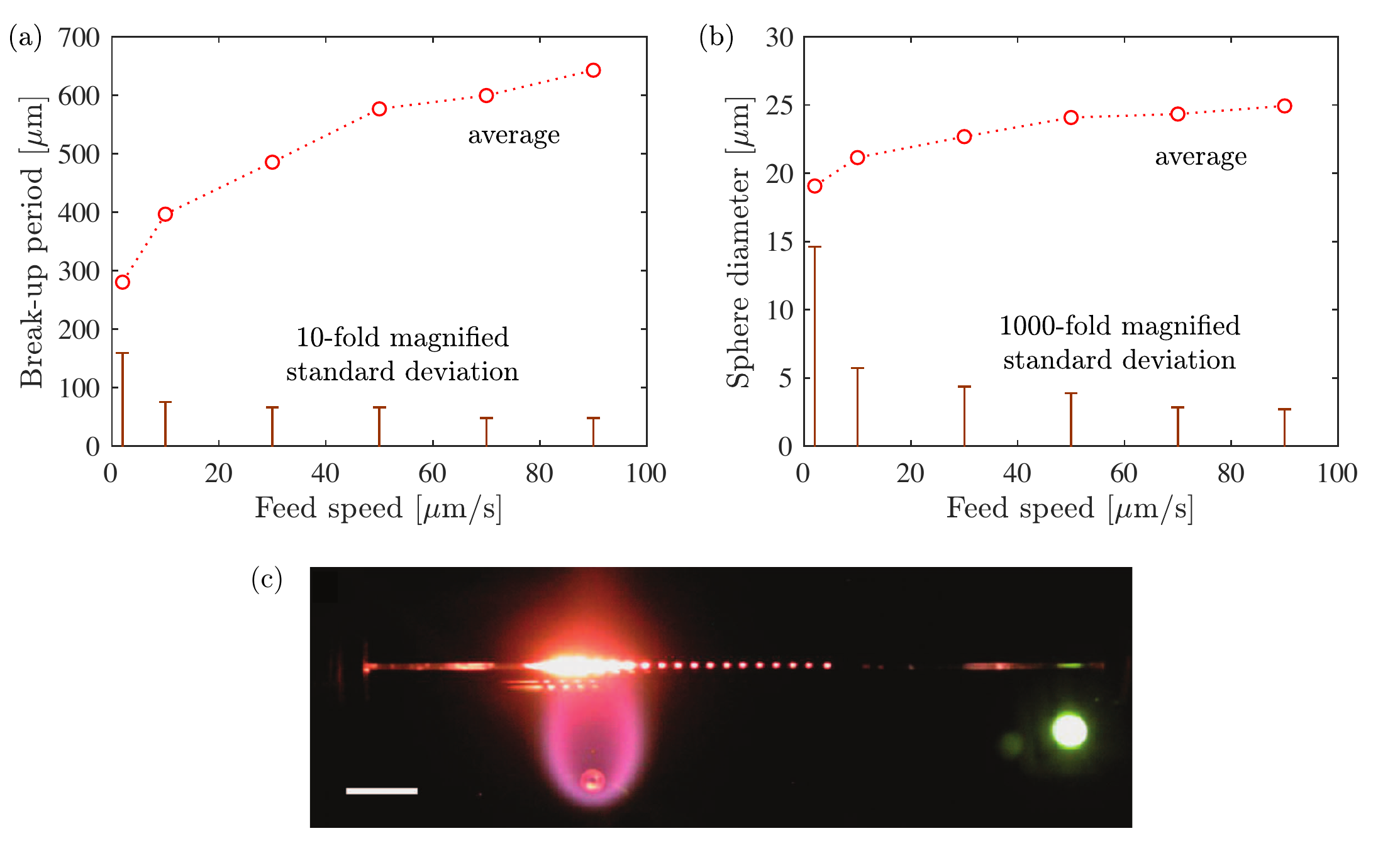}
\caption{Experimental results from Gumennik \textit{et al.}~\cite{gumennik2013} on the production of silicon particles by feeding a silicon-in-silica co-axial fiber with a $2 \, \mu\mathrm{m}$ core radius into a localized flame, triggering melting of the core and Rayleigh-Plateau instability of the silicon-silica interface. (a,b) The circles display the mean break-up period (a) and resulting sphere diameter (b) as a function of the feed speed. The bars show the standard deviation of the data, 10-fold magnified for the break-up period and 1000-fold magnified for the sphere diameter. (c) Photograph of a typical experiment, reproduced from Gumennik \textit{et al.}~\cite{gumennik2013}. The scale bar corresponds to $5 \,$mm.}
\label{fig:GumennikResults}
\end{figure}

For certain materials with very high viscosity contrast ratios, such as silicon-in-silica, the above method would produce large particles relatively to the size of the inner core, restricting the smallest attainable sphere diameter. In order to overcome this limitation, Gumennik \textit{et al.}~\cite{gumennik2013} developed a variant of the method, where instead of being uniformly heated, the fiber is fed at a given velocity into a spatially localized flame. In this way, melting of the inner silicon occurs locally and the formation of the spheres is dynamically coupled with the feed speed. Figure \ref{fig:GumennikResults}(a,b) reports the break-up period (a) and corresponding sphere diameter (b) that they obtained using silicon-in-silica fibers with a $2 \, \mu\mathrm{m}$ core radius and different feed speed values.
The circles show the average of the data while the bars show the 10-fold and 1000-fold magnified standard deviation of the break-up period and sphere diameter, respectively. Not only are the particle sizes reportedly smaller than achievable under an isothermal process, but there is also a clear and robust relationship between particle size and feed speed. The latter can therefore serve as a very convenient process parameter for adjusting the desired particle size, as opposed to tuning the temperature and/or material properties. Figure \ref{fig:GumennikResults}(c) shows a photograph of a typical experiment from Gumennik \textit{et al.}~\cite{gumennik2013}.

In this article, we try to rationalize the particle size observed in the experiments of Gumennik \textit{et al.}~\cite{gumennik2013} as well as its dependency on the feed speed. Such understanding of the dominant physical mechanisms at play in selecting the break-up wavelength would constitute a first step towards solving the inverse problem of determining the physical parameters and conditions required to obtain a desired particle size, which is essential to enable practical use of this fabrication technique. We will start by formulating a simple one-dimensional nonlinear governing equation for the motion of the silicon-silica interface, using long-wavelength approximations that have proven very accurate in the study of liquid jets \citep{eggers2008}. We will then employ linear stability analysis to try to elucidate the characteristic size of patterns that arise in this reduced governing equation. This approach is motivated by the similitude between the system under study, where spheres are formed at the tip of the molten silicon core, and the production of droplets at the tip of a microfluidic nozzle in a co-flowing ambient liquid \citep{cramer2004}. In the latter case, stability analysis tools have proven relevant at predicting the size of the droplets \citep{cordero2011}, although the effects of shear at the nozzle \citep{umbanhowar2000}, non-uniformity of the base flow \citep{augello2018}, and nonlinearity \citep{pier2001} are not entirely clear yet. As we will see later, however, linear stability analysis ultimately fails in our case. We therefore resort to a nonlinear stability analysis through numerical simulations of the reduced nonlinear governing equations, which recover, without any adjustable parameters, the relationship between sphere size and feed speed observed in Figure \ref{fig:GumennikResults}. This eventually shows that the formation of the silicon spheres is an intrinsically nonlinear process, in a way reminiscent of the dynamics of a slowly dripping faucet which has little to do with the instability of a hypothetical continuous jet solution.

The paper proceeds as follows. In Section \ref{sec:ProblemFormulation}, we describe the setup of the problem and derive a reduced one-dimensional model consisting of two coupled nonlinear differential equations governing the dynamics of the silicon-silica interface. Section \ref{sec:LinearStabilityAnalysis} then relates our unsuccessful attempts at predicting the particle size using linear stability analysis. Following this, we turn to numerical simulations of the nonlinear reduced model in Section \ref{sec:NonlinearStabilityAnalysis}, yielding good agreement with experimental results. Conclusions close the paper in Section \ref{sec:Conclusions}.

\section{Problem formulation}
\label{sec:ProblemFormulation}

\begin{figure}
\centering
\includegraphics[width=\textwidth]{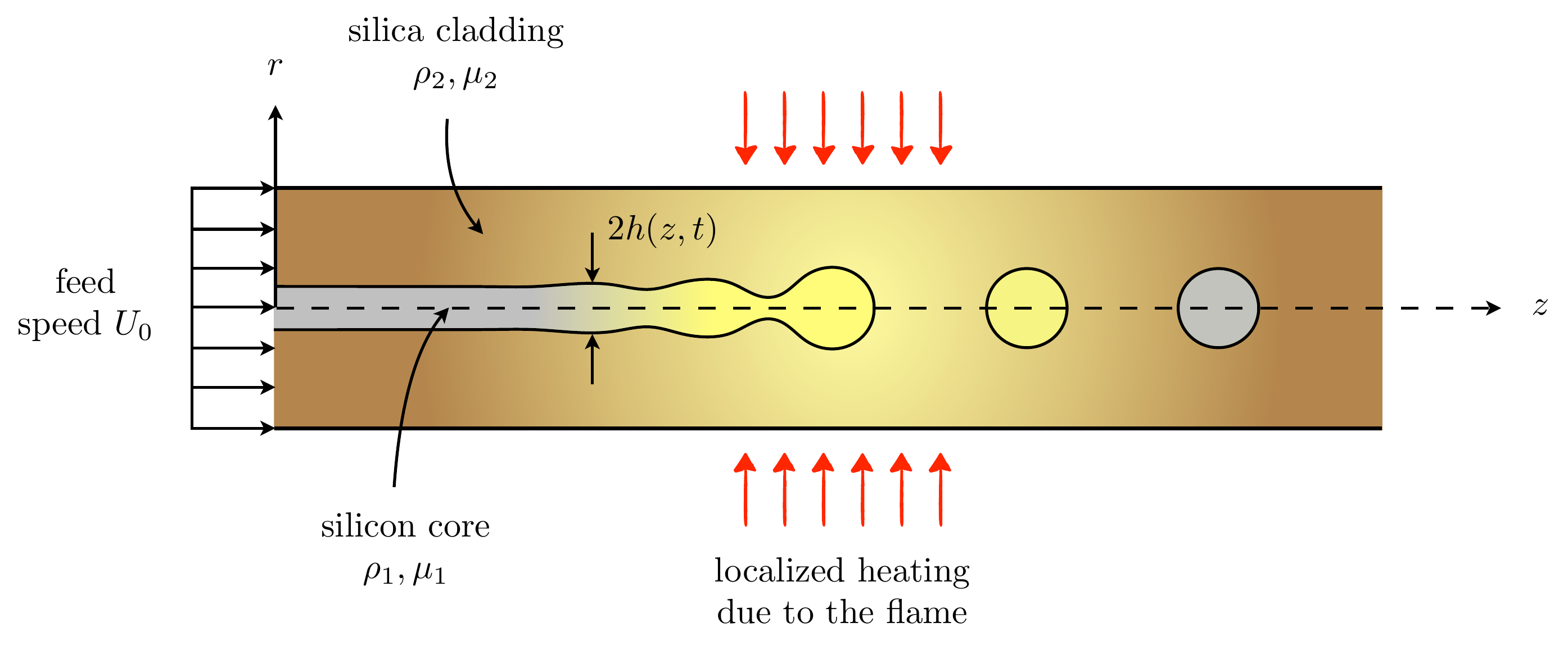}
\caption{Problem setup. A co-axial fiber consisting of a silicon core encased in a silica cladding is fed through a flame at a constant speed, causing the core to melt (pictured by the transition from gray to yellow color) while the cladding merely softens (pictured by the shift from darker to lighter brown color). Then, a capillary instability at the silicon-silica interface induces break-up of the silicon core into regular spheres, which re-solidify and remain trapped in the silica matrix upon exiting the flame. Note that the colors do not reflect the actual values of the viscosity.
}
\label{fig:CoaxialFiber}
\end{figure}

Let us consider the situation depicted in Figure \ref{fig:CoaxialFiber}, which reproduces the experimental setup of Gumennik \textit{et al.}~\cite{gumennik2013}.
A co-axial fiber made of a silicon core encased in a silica cladding is fed into a localized flame at a uniform velocity $U_0$. The local increase in temperature due to the flame causes the silicon core to melt while the silica cladding softens, at which point a capillary instability of the silicon-silica interface induces break-up of the core into regular silicon spheres. After the co-axial fiber leaves the flame, the silicon spheres re-solidify while cooling down and remain trapped within the silica matrix.

In this study, we focus on the instability mechanism leading to the formation of the spherical particles. We thus restrict our attention to the region where the silicon core is liquefied, which witnesses temperatures ranging from $T = 1414^\circ$C, the melting point of silicon, to $T \simeq 1760^\circ$C in the heart of the flame. Over this temperature range, the molten silicon core has relatively constant density $\rho_1 \simeq 2500 \, \mathrm{kg/m^3}$ and viscosity $\mu_1 \simeq 7 \cdot 10^{-4} \, \mathrm{Pa.s}$. By contrast, the silica cladding has similar density $\rho_2 \simeq \rho_1$ but much larger viscosity $\mu_2 \simeq 10^6 - 10^8 \, \mathrm{Pa.s}$, which varies by more than two orders of magnitude in this same temperature range. Therefore, the axial thermal gradient imposed by the flame gives rise to very strong spatial inhomogeneity in the system. Finally, the interfacial tension between silicon and silica is considered constant at $\gamma = 1.5 \, \mathrm{N/m}$ \cite{gumennik2013}.

We assume the flow to be axisymmetric and denote with $h(z,t)$ the position of the silicon/silica interface. Let $\mathbf{u}_i = u_i(r,z,t) \mathbf{e}_z + v_i(r,z,t) \mathbf{e}_r$ and $p_i(r,z,t)$ refer to the velocity and pressure fields in the molten silicon core $(i = 1)$ and outer silica $(i = 2)$. Before entering the flame, the system is uniformly advected at velocity $\mathbf{u}_0 = U_0 \mathbf{e}_z$ and the silicon core has constant radius $h_0 = 2 \, \mu\mathrm{m}$, defining the base state about which perturbations will grow after melting of the inner silicon. The outer radius of the co-axial fiber is $R = 140 \, \mu\mathrm{m}$ and is assumed to remain constant throughout the development of the instability.

\subsection{Equations of motion and boundary conditions}

The instability of the silicon-silica interface is driven by capillary forces and counteracted by inertial and viscous effects from both the silicon core and the silica fiber. Assuming for a moment that the silicon core is not affected by the outer silica, the time scale over which the instability is slowed down by inertia and viscosity would respectively be given by $\tau_{i,1} = (\rho_1 h_0^3/\gamma)^{1/2} \simeq 10^{-7} \, \mathrm{s}$ and $\tau_{v,1} = \mu_1 h_0/\gamma \simeq 10^{-9} \, \mathrm{s}$. The ratio of these time scales, called the Ohnesorge number $\mathit{Oh} = \tau_{v,1}/\tau_{i,1} \simeq 10^{-2}$, shows that viscous effects in the silicon are negligible compared with inertial effects. We therefore neglect the viscosity of the silicon and model the dynamics of the inner jet with the axisymmetric Euler equations,
\begin{subequations}
\begin{align}
\frac{\partial v_1}{\partial t} + v_1 \frac{\partial v_1}{\partial r} + u_1 \frac{\partial v_1}{\partial z} &= -\frac{1}{\rho_1} \frac{\partial p_1}{\partial r}, \label{eq:NSInner1} \\
\frac{\partial u_1}{\partial t} + v_1 \frac{\partial u_1}{\partial r} + u_1 \frac{\partial u_1}{\partial z} &= -\frac{1}{\rho_1} \frac{\partial p_1}{\partial z}. \label{eq:NSInner2}
\end{align}
\label{eq:NSInner}%
\end{subequations}
On the other hand, we describe the dynamics of the outer silica with the full axisymmetric Navier-Stokes equations,
\begin{subequations}
\begin{align}
\frac{\partial v_2}{\partial t} + v_2 \frac{\partial v_2}{\partial r} + u_2 \frac{\partial v_2}{\partial z} &= -\frac{1}{\rho_2} \frac{\partial p_2}{\partial r} + \nu_2 \left( \frac{\partial^2 v_2}{\partial r^2} + \frac{\partial^2 v_2}{\partial z^2} + \frac{1}{r} \frac{\partial v_2}{\partial r} - \frac{v_2}{r^2} \right), \label{eq:NSOuter1} \\
\frac{\partial u_2}{\partial t} + v_2 \frac{\partial u_2}{\partial r} + u_2 \frac{\partial u_2}{\partial z} &= -\frac{1}{\rho_2} \frac{\partial p_2}{\partial z} + \nu_2 \left( \frac{\partial^2 u_2}{\partial r^2} + \frac{\partial^2 u_2}{\partial z^2} + \frac{1}{r} \frac{\partial u_2}{\partial r}\right),
\end{align}
\label{eq:NSOuter}%
\end{subequations}
where $\nu_2 = \mu_2/\rho_2$. Since the outer radius of the fiber is two orders of magnitude larger than that of the silicon/silica interface, we consider the outer silica to be unbounded hence \eqref{eq:NSOuter} holds for $r > h(z,t)$ while \eqref{eq:NSInner} holds for $0 \le r < h(z,t)$. The continuity equation for both media reads
\begin{equation}
\frac{\partial v_i}{\partial r} + \frac{\partial u_i}{\partial z} + \frac{v_i}{r} = 0, \qquad i = 1,2.
\label{eq:Continuity}
\end{equation}
We are then left with the boundary conditions at the interface $r = h(z,t)$. The Laplace pressure due to surface tension imposes a discontinuity of the traction vector
\begin{equation}
\left. (\boldsymbol{\sigma}_1 - \boldsymbol{\sigma}_2) \mathbf{n} \right|_{r = h} = - \gamma \kappa \, \mathbf{n}.
\label{eq:DynamicBC}
\end{equation}
Here, $\kappa$ is the curvature of the interface,
\begin{equation}
\kappa = \frac{1}{h(1+h'^2)^{1/2}} - \frac{h''}{(1+h'^2)^{3/2}},
\label{eq:Curvature}
\end{equation}
with $h'$ and $h''$ denoting respectively the first and second derivatives of $h$ with respect to $z$, $\mathbf{n}$ is the outward normal to the interface,
\begin{equation}
\mathbf{n} = \frac{-h' \mathbf{e}_z + \mathbf{e}_r}{(1+h'^2)^{1/2}},
\end{equation}
and $\boldsymbol{\sigma}_1, \boldsymbol{\sigma}_2$ are respectively the stress tensors in the inner and outer fluids,
\begin{gather}
\boldsymbol{\sigma}_1 = -p_1 \boldsymbol{I}, \\
\boldsymbol{\sigma}_2 = -p_2 \boldsymbol{I} + \mu_2 (\nabla \mathbf{u}_2 + \nabla \mathbf{u}_2^\mathsf{T}).
\end{gather}
The projection of the stress condition \eqref{eq:DynamicBC} along the normal direction gives\begin{equation}
\left. p_1 - p_2 + \frac{2 \mu_2}{1+h'^2} \left[ \frac{\partial v_2}{\partial r} + \frac{\partial u_2}{\partial z} h'^2 - \left( \frac{\partial u_2}{\partial r} + \frac{\partial v_2}{\partial z} \right) h' \right] \right|_{r=h} = \gamma \kappa.
\label{eq:NormalStressBC}
\end{equation}
The second boundary condition comes from continuity of the normal velocity of the interface with that of the two fluids
\begin{equation}
\left. \frac{\partial h}{\partial t} + u_i \frac{\partial h}{\partial z} = v_i \right|_{r = h}, \qquad i = 1,2,
\label{eq:KinematicBC}
\end{equation}
which also ensures continuity of the normal velocity in the fluid across the interface.

\subsection{Inner silicon core}

The dynamics of the inner silicon jet can be simplified using a long-wavelength approximation that reduces the axisymmetric system to a one-dimensional equation \citep{eggers1994,eggers2008}. Exploiting the fact that the radial length scale $h_0$ of the jet is much smaller than its axial length scale $\lambda \sim 1/k$, where $k$ is a typical interface deformation wavenumber, the velocity and pressure fields can be expanded in Taylor series with respect to $r$
\begin{subequations}
\begin{align}
u_1(r,z,t) &= \bar{u}_{10}(z,t) + \bar{u}_{12}(z,t)r^2 + \dots, \\
v_1(r,z,t) &= -\frac{1}{2}\bar{u}_{10}'(z,t)r - \frac{1}{4}\bar{u}_{12}'(z,t)r^3 + \dots, \\
p_1(r,z,t) &= \bar{p}_{10}(z,t) + \bar{p}_{12}(z,t) r^2 + \dots,
\end{align}
\end{subequations}
where $v_1$ is chosen to enforce incompressibility of the velocity field. Inserting these expansions into the axisymmetric Euler equations \eqref{eq:NSInner1} or \eqref{eq:NSInner2} and solving at leading order gives
\begin{equation}
\frac{\partial \bar{u}_{10}}{\partial t} + \bar{u}_{10} \frac{\partial \bar{u}_{10}}{\partial z} = - \frac{1}{\rho_1} \frac{\partial \bar{p}_{10}}{\partial z},
\label{eq:OneDimensionalEquation1}
\end{equation}
while the kinematic condition \eqref{eq:KinematicBC} gives at lowest order
\begin{equation}
\frac{\partial h}{\partial t} + \bar{u}_{10} \frac{\partial h}{\partial z} = - \frac{1}{2} \frac{\partial \bar{u}_{10}}{\partial z} h.
\label{eq:OneDimensionalEquation2}
\end{equation}
These are a set of coupled one-dimensional equations for the leading-order inner fluid velocity $\bar{u}_{10}$ and the interface position $h$. The pressure $\bar{p}_{10}$, which couples the dynamics of the inner silicon core with the outer silica through the normal stress boundary condition \eqref{eq:NormalStressBC}, remains unknown at this point.

\subsection{Outer silica cladding}
\label{sec:OuterSilicaCladding}

For the outer silica, separation of scales again enables us to simplify the governing equations. Since the outer radius $R$ of the fiber is much larger than its axial length scale $\lambda \sim 1/k$, where $k$ is a typical interface deformation wavenumber, we neglect variations of the axial velocity and suppose that it remains equal to its base flow value $u_2 = U_0$. In this way, we assume that perturbations to the interface position only generate a purely radial, expanding or contracting perturbed velocity field $v_2(r,z,t)$. Furthermore, we will only retain terms with a linear contribution in the perturbation, with the exception of the interface curvature $\kappa$. Under these assumptions, the continuity equation \eqref{eq:Continuity} becomes
\begin{equation}
\frac{1}{r} \frac{\partial (r v_2)}{\partial r} = 0,
\label{eq:ContinuityOuter}
\end{equation}
and the normal stress boundary condition \eqref{eq:NormalStressBC} reduces to
\begin{equation}
\left. \bar{p}_{10} - p_2 + 2 \mu_2 \frac{\partial v_2}{\partial r} \right|_{r=h} = \gamma \kappa. 
\label{eq:ReducedNormalStressBC}
\end{equation}
The kinematic boundary condition \eqref{eq:KinematicBC} at the interface,
\begin{equation}
\left. \frac{\partial h}{\partial t} + U_0 \frac{\partial h}{\partial z} = v_2 \right|_{r = h},
\end{equation}
can be combined with the continuity equation \eqref{eq:ContinuityOuter}, integrated in the radial direction, to give an explicit expression for $v_2$ in terms of the interface deformation,
\begin{equation}
v_2 = \frac{h}{r} \left( \frac{\partial h}{\partial t} + U_0 \frac{\partial h}{\partial z} \right).
\label{eq:OuterVelocity}
\end{equation}
We now make the assumption that the pressure $p_2$ in the outer silica is approximately constant. This assumption is justified in Appendix \ref{app_0}, where we show that solving for $p_2$ using the momentum equation \eqref{eq:NSOuter1} ultimately leads to a dispersion relation that is virtually indistinguishable from that obtained by neglecting $p_2$. Inserting the above expression for $v_2$ into the normal stress condition \eqref{eq:ReducedNormalStressBC} and setting $p_2 = \mathrm{cst}$ yields an expression for the leading-order inner pressure,
\begin{equation}
\bar{p}_{10} =  \gamma  \kappa + \frac{2 \mu_2}{h} \left( \frac{\partial h}{\partial t} + U_0 \frac{\partial h}{\partial z} \right) + \mathrm{cst},
\label{eq:InnerPressure}
\end{equation}
where the first term is the Laplace pressure jump at the interface, and the second term is the normal component of the viscous stress in the outer silica at the interface.

\subsection{Reduced nonlinear governing equations}

As a final step, we insert expression \eqref{eq:InnerPressure} for $\bar{p}_{10}$ into the one-dimensional equation \eqref{eq:OneDimensionalEquation1} describing the dynamics of the inner jet. Combined with \eqref{eq:OneDimensionalEquation2}, we arrive at a coupled system of two nonlinear governing equations for the leading-order inner velocity $\bar{u}_{10}$ and interface radius $h$,
\begin{subequations}
\begin{align}
\frac{\partial \bar{u}_{10}}{\partial t} + \bar{u}_{10} \frac{\partial \bar{u}_{10}}{\partial z} &= - \frac{\gamma}{\rho_1} \frac{\partial \kappa}{\partial z} - \frac{2}{\rho_1} \frac{\partial}{\partial z} \left[ \frac{\mu_2}{h} \left( \frac{\partial h}{\partial t} + U_0 \frac{\partial h}{\partial z} \right) \right], \\
\frac{\partial h}{\partial t} + \bar{u}_{10} \frac{\partial h}{\partial z} &= - \frac{1}{2} \frac{\partial \bar{u}_{10}}{\partial z} h,\end{align}
\label{eq:GoverningEquations}%
\end{subequations}%
with the interface curvature expressed as
\begin{equation}
\kappa = \frac{1}{h(1+h'^2)^{1/2}} - \frac{h''}{(1+h'^2)^{3/2}}.
\label{eq:Curvature2}
\end{equation}
We remind the reader that $\mu_2(z)$ is a strongly varying function of $z$. These two governing equations constitute a reduced nonlinear model for the motion of the interface in the silicon-in-silica fiber, and form the starting point of the subsequent analysis.

\section{Linear stability analysis}
\label{sec:LinearStabilityAnalysis}

In this section, we try to rationalize the droplet size experimentally observed by Gumennik \textit{et al.}~\cite{gumennik2013} using linear stability analysis, which has been successful at elucidating the characteristic size of patterns arising from a wide range of interfacial instabilities (for a review, see \cite{gallaire2017}). 
Although the system under study is non-homogeneous due to the strong axial dependency of the silica viscosity, we perform the stability analysis in a local framework wherein the system is considered uniform at each axial location.

\subsection{Dispersion relation}
\label{sec:DispersionRelation}

We begin by deriving the dispersion relation describing the local instability characteristics of the system defined by the coupled set of equations \eqref{eq:GoverningEquations}. This is done by setting the viscosity of the outer silica to be constant and equal to $\mu_2(z^*)$, where $z^*$ is the axial location of interest. The system \eqref{eq:GoverningEquations} is then axially uniform and one can find the dispersion relation governing the growth of small perturbations to $(h,\bar{u}_{10})$ about the base state $(h_0,U_0)$ by considering the normal mode expansion
\begin{subequations}
\begin{align}
h(z,t) &= h_0 + \epsilon a e^{i(kz-\omega t)}, \\
\bar{u}_{10}(z,t) &= U_0 + \epsilon b e^{i(kz-\omega t)},
\end{align}
\end{subequations}
where $\epsilon \ll 1$, $k$ and $\omega$ are respectively the perturbation wavenumber and frequency, which may both be complex, and $a$ and $b$ are complex constants. Inserting the above expansion into equations \eqref{eq:GoverningEquations} and linearizing about $(h_0,U_0)$ leads to the dispersion relation
\begin{equation}
\frac{\rho_1 h_0^3}{\gamma} (\omega - U_0 k)^2 + i \frac{\mu_2 h_0}{\gamma} (kh_0)^2 (\omega - U_0 k) + \frac{1}{2} \left[ (kh_0)^2-(kh_0)^4 \right] = 0.
\label{eq:DispersionRelation}
\end{equation}
Interestingly, the above dispersion relation is identical to that obtained by Eggers and Dupont \cite{eggers1994} for a jet with density $\rho = \rho_1$ and viscosity $\mu = \mu_2/3$ in an inert medium. Although both dispersion relations are obtained using the same long-wavelength approximation, the similarity is nonetheless surprising given the different forms and origins of the viscous term appearing in the reduced governing equations.

Before discussing wavelength selection, let us first investigate a possible simplification of the dispersion relation. Equation \eqref{eq:DispersionRelation} shows that disturbances are driven by surface tension (third term) and simultaneously slowed down by inertia from the inner silicon (first term) and by viscous forces from the outer silica (second term). The time scale associated with the inertial term is on the order of $\tau_{i,1} = (\rho_1 h_0^3/\gamma)^{1/2} \simeq 10^{-7} \, \mathrm{s}$ while its viscous counterpart is in the range $\tau_{v,2} = \mu_2 h_0/\gamma \simeq 1 $--$ 10^2 \, \mathrm{s}$ depending on the local temperature of the system. The ratio of these two time scales defines a mixed Ohnesorge number $\mathit{Oh}' = \tau_{v,2}/\tau_{i,1} \simeq 10^7$--$10^9 \gg 1$, which suggests that inertial effects are negligible. We are therefore tempted to set 
$\rho_1 = 0$, leading to the dispersion relation
\begin{equation}
\omega = U_0 k + i\frac{\gamma}{2 \mu_2 h_0}[1-(kh_0)^2],
\label{eq:ViscousDispersionRelation}
\end{equation}
which represents the purely viscous limit of \eqref{eq:DispersionRelation}.

\subsection{Temporal stability}
\label{sec:LocalTemporalStability}

\begin{figure}
\centering
\includegraphics[width=\textwidth]{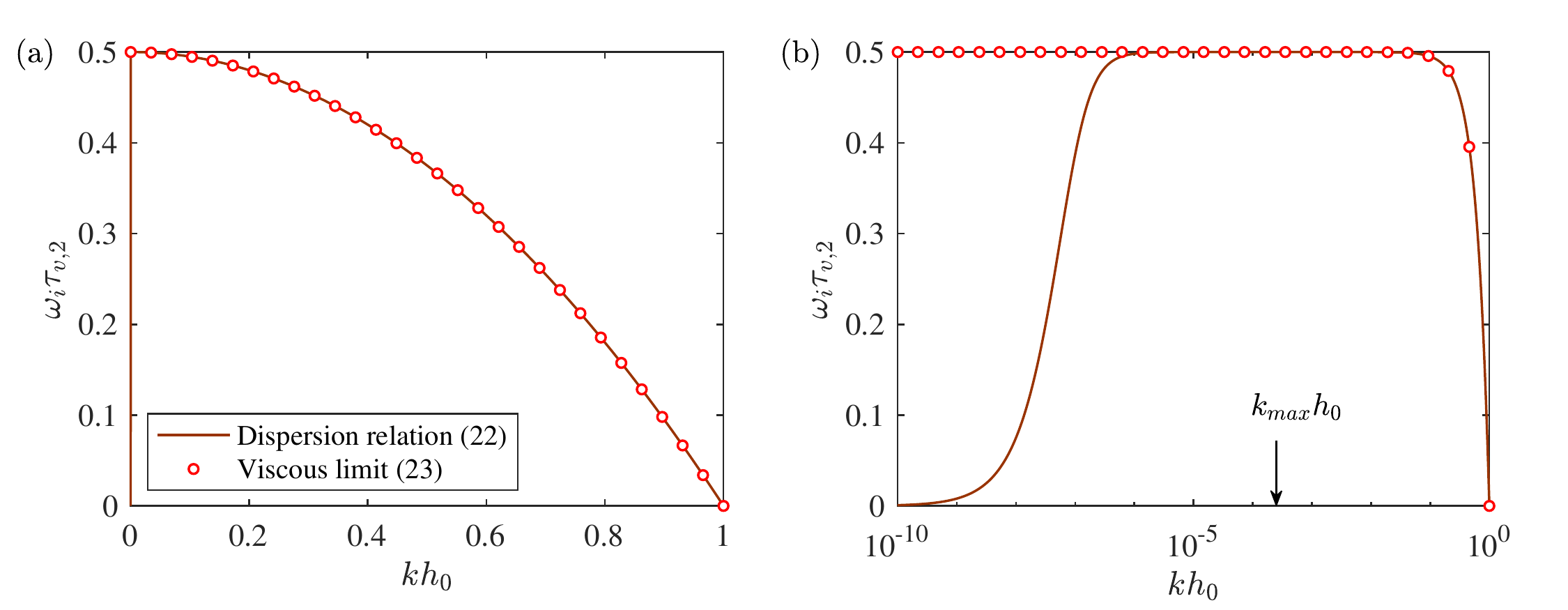}
\caption{Temporal growth rate $\omega_i$ as a function of the real wavenumber $k$ from the dispersion relation \eqref{eq:DispersionRelation} and its viscous limit \eqref{eq:ViscousDispersionRelation}, in (a) linear and (b) logarithmic wavenumber scales. The most unstable wavenumber predicted by \eqref{eq:DispersionRelation} is  $k_{max} h_0 \simeq 2.40 \cdot 10^{-4}$, while for \eqref{eq:ViscousDispersionRelation} it is $k_{max} = 0$.}
\label{fig:TemporalDispersionRelation}
\end{figure}

First, we compare the dispersion relations \eqref{eq:DispersionRelation} and \eqref{eq:ViscousDispersionRelation} on the basis of their temporal stability predictions. The temporal growth rate of perturbations is given by $\omega_i$, the imaginary part of $\omega$, for real values of $k$. Figure \ref{fig:TemporalDispersionRelation} presents $\omega_i$ -- nondimensionalized with the viscous time scale $\tau_{v,2} = \mu_2 h_0/\gamma$ -- as a function of the dimensionless wavenumber $kh_0$, for $\mu_2 = 10^6 \, \mathrm{Pa.s}$ and all other parameters as given in Section \ref{sec:ProblemFormulation}.
This value of $\mu_2$ is representative of the heart of the flame, where the silica viscosity is lowest hence the interface most unstable. It is therefore not unreasonable to assume that this region will play the largest role in setting the length scale of the resulting spheres. Results are shown for the dispersion relation \eqref{eq:DispersionRelation} together with its viscous limit \eqref{eq:ViscousDispersionRelation}, in linear (a) and logarithmic (b) wavenumber scale. Even though the agreement between the two dispersion relations is excellent for $kh_0 > 10^{-5}$, the viscous limit \eqref{eq:ViscousDispersionRelation} predicts that the most amplified wavenumber is $k_{max} = 0$. This makes it ill-posed, since such a wavenumber would correspond to an infinite disturbance wavelength. On the other hand, \eqref{eq:DispersionRelation} predicts a maximum growth rate at $k_{max} h_0 \simeq 2.40 \cdot 10^{-4}$.

We now investigate the implications of these results for wavelength selection, which in the temporal framework is dictated by the temporally most unstable wavenumber $k_{max}$. Using $h_0 = 2 \, \mu\mathrm{m}$, the value of $k_{max}$ predicted by \eqref{eq:DispersionRelation} corresponds to a wavelength $\lambda_m = 2\pi/k_{max} \simeq 52.4 \, \mathrm{mm}$ -- two orders of magnitude larger than the break-up period reported by Gumennik \textit{et al.}~\cite{gumennik2013} over a range of advection velocities $U_0$, see Figure \ref{fig:GumennikResults}. Furthermore, the temporal stability predictions for the instability wavelength selected by the system are also insensitive to the advection velocity $U_0$, contrary to the observations reported in Figure \ref{fig:GumennikResults}. 

\subsection{Spatio-temporal stability}

We now turn to a spatio-temporal stability analysis, which generalizes the previous temporal analysis by taking into account the effect of the advection velocity $U_0$ of the system on its stability properties and selected perturbation wavelength. In this framework, one characterizes the impulse response of the system to a localized perturbation, which generates a coherent wave packet that will grow in time and space as long as the system is temporally unstable. The asymptotic spatio-temporal behavior of this wave packet in the laboratory frame will naturally depend on the advection velocity of the system, and can be described in terms of an absolute wavenumber $k_0$ and absolute frequency $\omega_0$. These are defined by the following saddle point condition together with the dispersion relation \citep{huerre1990}
\begin{equation}
\frac{\mathrm{d} \omega}{\mathrm{d} k}(k_0) = 0, \quad \omega_0 = \omega(k_0),
\label{eq:ZeroGroupVelocityCondition}
\end{equation}
where both $k_0$ and $\omega_0$ are allowed to be complex. 
The imaginary part $\omega_{0i}$ of the absolute frequency $\omega_0$ characterizes the temporal evolution of the impulse response wave packet observed at a fixed spatial location. Its sign therefore determines the spatio-temporal instability behavior of the system in the laboratory frame. If $\omega_{0i} > 0$, then the system is absolutely unstable -- localized perturbations grow fast enough to overcome system advection and eventually invade the entire domain. If $\omega_{0i} < 0$, then the system is convectively unstable -- localized perturbations are convected away before they are able to grow in the laboratory frame.

Here, we calculate the absolute wavenumber $k_0$ and frequency $\omega_0$ of the silicon-in-silica fiber using Bers' pinch point condition \cite{bers1983}, an equivalent set of equations to \eqref{eq:ZeroGroupVelocityCondition} that avoids the need to express $\omega$ as a function of $k$, and takes the form 
\begin{equation}
\frac{\partial \Delta}{\partial k}(k_0,\omega_0) = 0, \quad \Delta(k_0,\omega_0) = 0,
\label{eq:PinchPointCondition}
\end{equation}
where $\Delta(k,\omega) = 0$ is the local dispersion relation of the system. We apply the above pinch point condition to the dispersion relation \eqref{eq:DispersionRelation}. First, we define the dimensionless frequency $\tilde{\omega} = \omega \tau_{v,2}$ and wavenumber $\tilde{k} = k h_0$, so that  \eqref{eq:DispersionRelation} becomes, in nondimensional form,
\begin{equation}
\Delta(\tilde{k},\tilde{\omega}) = \frac{1}{\mathit{Oh}'^2} (\tilde{\omega} - \mathit{Ca} \tilde{k})^2 + i \tilde{k}^2 (\tilde{\omega} - \mathit{Ca} \tilde{k}) + \frac{1}{2} ( \tilde{k}^2-\tilde{k}^4 ) = 0,
\label{eq:DispersionRelationND}
\end{equation}
with $\mathit{Oh}' = \mu_2/\sqrt{\rho_1 \gamma h_0}$ the mixed Ohnesorge number defined in Section \ref{sec:DispersionRelation}, and $\mathit{Ca} = \mu_2 U_0/\gamma$ the capillary number. Then, the first equation in condition \eqref{eq:PinchPointCondition} directly follows as
\begin{equation}
\frac{\partial \Delta}{\partial \tilde{k}}(\tilde{k},\tilde{\omega}) = - 2\frac{\mathit{Ca}}{\mathit{Oh}'^2} (\tilde{\omega}-\mathit{Ca} \tilde{k}) + 2i \tilde{k} (\tilde{\omega}-\mathit{Ca} \tilde{k}) - i \mathit{Ca} \tilde{k}^2 + (\tilde{k}-2\tilde{k}^3) = 0.
\label{eq:PinchPointConditionFirstEquation}
\end{equation}
As before, the results that we will obtain for given values of $\mathit{Oh}'$ and $\mathit{Ca}$ must be interpreted locally, in the sense that they relate to specific axial stations in the system. The axial dependency of the silica viscosity $\mu_2$ imparts an axial variation to both $\mathit{Oh}'$ and $\mathit{Ca}$. $\mathit{Oh}'$ decreases from $10^9$ to about $10^7$ as the fiber enters the flame, independently of the feed speed $U_0$. The latter, however, affects the range of values of $\mathit{Ca}$. For feed velocity $U_0 = 1 \, \mu\mathrm{m}/\mathrm{s}$, $\mathit{Ca}$ decreases from $10^2$ to about $1$, while for high feed velocity $U_0 = 100 \, \mu\mathrm{m}/\mathrm{s}$, $\mathit{Ca}$ correspondingly decreases from $10^4$ to about $10^2$.

\begin{figure}[t]
\centering
\includegraphics[width=\textwidth]{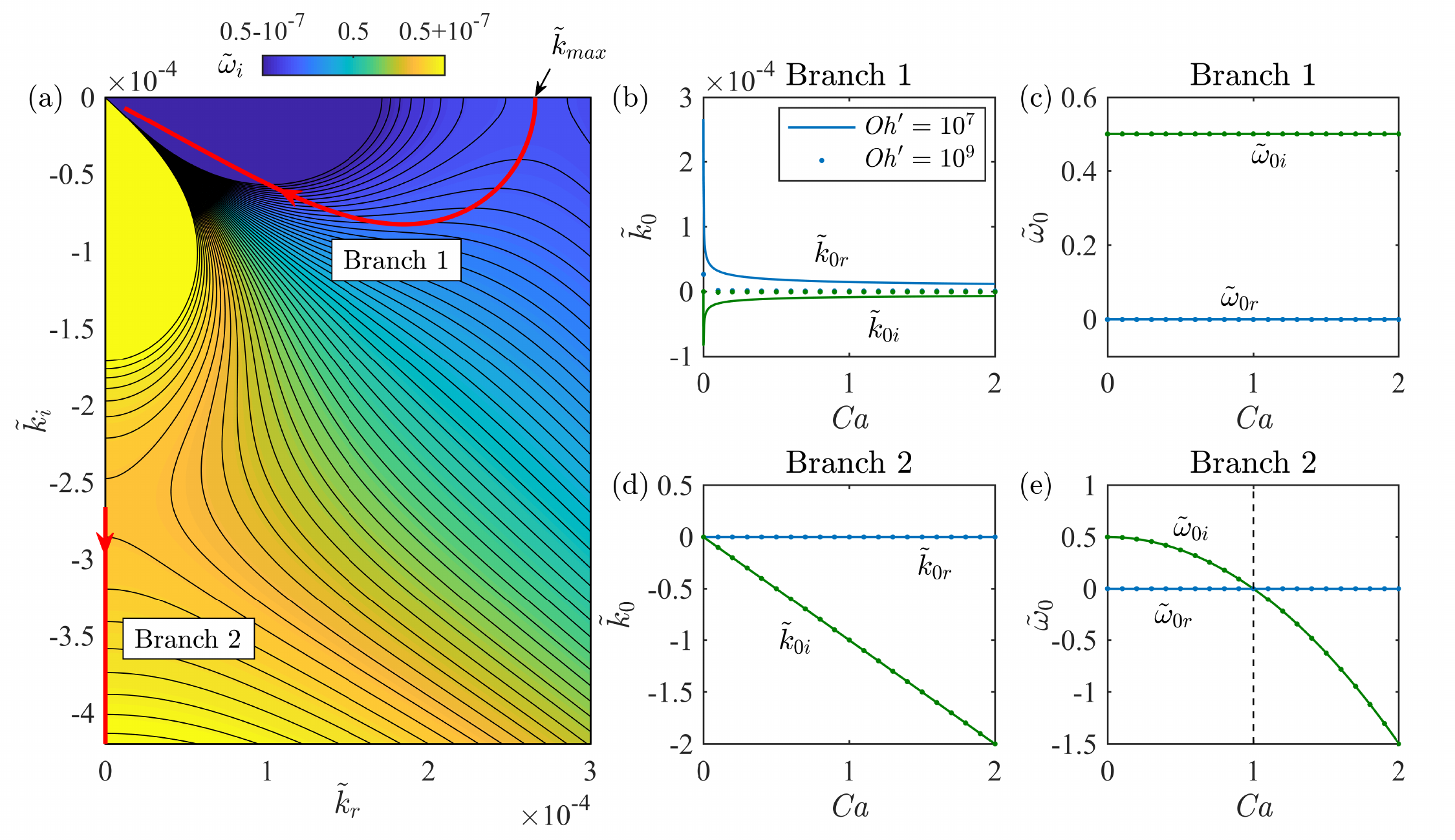}
\caption{Spatio-temporal stability properties of the dispersion relation \eqref{eq:DispersionRelation}. (a) Level curves of $\tilde{\omega}_i$ as a function of complex $\tilde{k}$ for $\mathit{Ca} = 0$ and $\mathit{Oh}' = 10^7$, given by \eqref{eq:DispersionRelationND}. The red lines trace out the locus of absolute wavenumbers $\tilde{k}_0$ as $\mathit{Ca}$ is increased from $0$ to $2$, which is obtained from the coupled system of equations \eqref{eq:DispersionRelationND} and \eqref{eq:PinchPointConditionFirstEquation}. There are two solution branches, denoted here branch 1 and branch 2. (b)--(e) Absolute wavenumber $\tilde{k}_0$ and absolute frequency $\tilde{\omega}_0$ pertaining to these two solution branches as a function of $\mathit{Ca}$, for $\mathit{Oh}' = 10^7$ (solid line) and $\mathit{Oh}' = 10^9$ (dots).}
\label{fig:SaddlePoints}
\end{figure}

We solve the coupled system of equations \eqref{eq:DispersionRelationND} and \eqref{eq:PinchPointConditionFirstEquation} for $\mathit{Oh}' = 10^7$ and various values of $\mathit{Ca}$ using a Newton-Raphson iterative scheme with tolerance $10^{-15}$ on the $L_2$ norm of the residual. For each value of $\mathit{Ca}$, we find that there are two absolute wavenumber and absolute frequency pairs $(\tilde{k}_0,\tilde{\omega}_0)$ that solve the pinch point condition. These two solution branches are shown in Figure \ref{fig:SaddlePoints}(a) by the red lines labelled branch 1 and branch 2, which trace out (in the direction of the arrow) the locus of absolute wavenumbers $\tilde{k}_0$ in the complex $\tilde{k}$-plane as $\mathit{Ca}$ is increased from 0 to 2.
In the same figure, we display for $\mathit{Ca} = 0$ the contour levels of $\tilde{\omega}_i(\tilde{k})$, the imaginary part of $\tilde{\omega}$ obtained by solving the dispersion relation \eqref{eq:DispersionRelationND} for complex values of $\tilde{k}$. In accordance with the saddle point condition \eqref{eq:ZeroGroupVelocityCondition}, which states that $\tilde{k}_0$ is a saddle point of $\tilde{\omega}_i(\tilde{k})$, we observe that the start points of both red curves coincide with a saddle point of the dispersion relation. In the case of branch 1, this saddle point actually corresponds to the temporally most unstable wavenumber $k_{max}$ identified in the previous section.

Figures \ref{fig:SaddlePoints}(b)--(e) display the absolute wavenumber $\tilde{k}_0$ and absolute frequency $\tilde{\omega}_0$ pertaining to branches 1 and 2 as a function of $\mathit{Ca}$, for $\mathit{Oh}' = 10^7$ and $10^9$. The convective or absolute instability behavior of the system for specific values of $\mathit{Oh}'$ and $\mathit{Ca}$ is given by the sign of $\tilde{\omega}_{0i}$, the imaginary part of the absolute frequency $\tilde{\omega}_0$. In our case, however, there are two solution branches that lead to different characterizations. According to branch 1, the system remains absolutely unstable for $\mathit{Ca}$ between 0 and 2 since $\tilde{\omega}_{0i}$ is always positive. By contrast, branch 2 indicates a transition from absolute to convective instability with $\tilde{\omega}_{0i}$ turning negative at $\mathit{Ca} = 1$. 
Since the local spatio-temporal instability behavior of the system is generally dictated by the saddle point with highest $\tilde{\omega}_{0i}$, Figure \ref{fig:SaddlePoints} suggests that branch 1 is the most relevant one for all finite values of $\mathit{Ca}$. 

In the context of pattern formation, the distinction between absolute and convective instability is crucial for wavelength selection \citep{duprat2007,gallaire2017}, even in systems with streamwise-varying properties. Flows which locally undergo a transition from convective to absolute instability at some downstream station exhibit a saturated pattern with a well-defined wavelength given by $2\pi U_0/\omega_{0r}$, where $\omega_{0r}$ is the real part of the local absolute frequency $\omega_0$ at the upstream boundary of the absolute instability region \citep{pier1998,pier2001b}. Conversely, flows which are convectively unstable everywhere amplify incoming disturbances as the latter travel downstream, resulting in a broader distribution of pattern wavelengths. Returning to our system, we observe that the absolute frequency $\tilde{\omega}_0$ corresponding to branch 1 is virtually unchanged as $\mathit{Oh}'$ and $\mathit{Ca}$ vary, with its real and imaginary parts $\tilde{\omega}_{0r}$ and $\tilde{\omega}_{0i}$ being equal to $0$ and $0.5$, respectively. We have verified that this remains true for values of $\mathit{Ca}$ as large as $10^4$. This implies that within the operating conditions of the experiments of Gumennik \textit{et al.}~\cite{gumennik2013}, the instability is everywhere locally absolute in the region where the inner silicon is liquefied. As such, the dominant wavelength selected by the system, which is in principle determined by $2\pi U_0/\omega_{0r}$, is predicted to be infinite for all values of the feed speed $U_0$. In conclusion, it becomes clear that the behavior of small interface perturbations -- governed by linear stability analysis -- is irrelevant to the length scale of the resulting silicon spheres.

\section{Nonlinear stability analysis}
\label{sec:NonlinearStabilityAnalysis}

In view of the failure of linear stability analysis at predicting the break-up wavelength selected by the system, we hypothesize that nonlinear effects play a predominant role and we turn in this section to numerical simulations of the nonlinear governing equations \eqref{eq:GoverningEquations}. From here on, the inlet will refer to the melting location of the silicon core, which happens when its temperature increases above $T = 1414^\circ$C, the melting point of silicon.

\subsection{Dimensionless governing equations}

In order to nondimensionalize the governing equations \eqref{eq:GoverningEquations}, we select the silicon core inlet radius $h_0$ as the characteristic length scale and the feed speed $U_0$ as the characteristic velocity scale. We denote $v = \bar{u}_{10}/U_0$ the dimensionless velocity, $\tilde{z} = z/h_0$ the dimensionless axial coordinate, and $\tilde{t} = t/(h_0/U_0)$ the dimensionless time. Additionally, in order to remove the singularity in expression \eqref{eq:Curvature2} for the curvature, we describe the interface radius in terms of the dimensionless function $f = (h/h_0)^2$. Then, the inlet conditions translate as $f(\tilde{z} = 0, \tilde{t})=1$ and $v(\tilde{z} = 0,\tilde{t})=1$. The governing equations \eqref{eq:GoverningEquations} become
\begin{subequations}\label{eqN2}
\begin{align}
\mathit{We} \left(\frac{\partial v}{\partial \tilde{t}} + v\frac{\partial v}{\partial \tilde{z}}\right ) &=  - \frac{\partial \tilde{\kappa}}{\partial \tilde{z}} - \frac{\partial}{\partial \tilde{z}} \left[ \frac{\mathit{\mathit{Ca}}_{\tilde{z}}}{f} \left(-\frac{\partial (fv)}{\partial \tilde{z}} + \frac{\partial f}{\partial \tilde{z}}\right) \right],
\\
\frac{\partial f}{\partial \tilde{t}} &= -\frac{\partial (fv)}{\partial \tilde{z}},
\\ 
\tilde{\kappa} &= \frac{(2-f'')f + f'^2}{2 ( f'^2/4 + f)^{3/2}},
\end{align}
\end{subequations}
where $\mathit{We}$ and $\mathit{\mathit{Ca}}_{\tilde{z}}$ are respectively the Weber and axially-dependent capillary numbers. Here, the Weber number, expressed as $\mathit{We} = {\rho_1 h_0 {U_0}^2}/{\gamma}$, measures the relative importance of the kinetic energy of the silicon core with respect to the silicon-silica interfacial energy. The axially-dependent capillary number, expressed as $\mathit{Ca}_{\tilde{z}} = {\mu_2(\tilde{z}) U_0}/{\gamma}$, compares the viscous force due to the spatially-varying outer silica viscosity with the silicon-silica surface tension force. We now proceed, in the next section, to the description of the numerical scheme used for solving \eqref{eqN2}.
\subsection{Numerical scheme} \label{NumScheme}
\label{NumericalScheme}
The governing equations \eqref{eqN2} are first discretized in space, after which the resulting ODEs are integrated in time. Diffusion terms are evaluated using second-order finite differences, with a central scheme for intermediate nodes and a forward or backward scheme for boundary nodes. Advection terms are obtained using a weighted upwind scheme inspired by Spalding's hybrid difference scheme \cite{spalding1972novel}. Unlike the latter, which approximates the convective derivative using a combination of central and upwind schemes, we evaluate the derivative based on a combination of forward and backward finite differences. An advection term $da/dz$ is evaluated at node $i$ as
\begin{equation}\label{eqNb1}
\left( \frac{d a}{d z} \right)_i = \beta \left( \frac{d a}{d z} \right)_{i,b} + (1-\beta) \left( \frac{d a}{d z} \right)_{i,f},
\end{equation}
where indices $b$ and $f$ refer to the backward and forward finite difference schemes, and $\beta$ is a weight coefficient that depends on the local value of velocity $v$ at node $i$ together with a parameter $\alpha$,
\begin{equation}\label{eqNb2}
\beta = \frac{\tanh(\alpha v_i)+1}{2}.
\end{equation}
For the range of feed velocities considered in this study, numerical stability was always ensured by using a 10-point stencil. Thus, the backward difference term relies on a stencil that spans nodes $i-5$ to $i+4$, and the forward difference term employs nodes $i-4$ to $i+5$. For large enough downstream or upstream velocities, $\beta$ will tend to $1$ or $0$ respectively; hence \eqref{eqNb1} reduces to a regular upwind difference scheme. For smaller velocity magnitudes in between, \eqref{eqNb1} produces a weighted combination of backward and forward differences. In our simulations, we choose $\alpha = 50$ so that the transition between the backward and forward difference schemes mostly occurs when $|v| < 0.05$. Finally, advection terms at nodes close to the boundary are evaluated based on the values of the closest 9 adjoining nodes. 

After obtaining all spatial derivatives, the resulting ODEs are integrated using the MATLAB solver \texttt{ode23tb}, which implements a trapezoidal rule and backward differentiation formula known as TR-BDF2 \citep{bank1985transient}, and uses a variable time step to reduce the overall simulation time. The jet interface is initialized as a cylinder of constant radius (equal to the inner core inlet radius $h_0$) and constant velocity (equal to feed speed $U_0$), that is $f(z,0) = 1$ and $v(z,0) = 1$. The boundary conditions at the inlet are defined as $f(0,t)=1$ and $v(0,t)=1$. No boundary conditions are defined at $z=L$, where $L$ is the size of the spatial domain. 

At every time step, the solution is evaluated for three conditions: (\textit{i}) \textit{Pinch-off (break-up):} It is defined as when the value of $f$ passes below a threshold value of $10^{-5}$. The corresponding time $T_{po}$ is saved and the position of the jet tip is updated as $N_{tip} = N_{po}$, where $N_{po}$ is the pinch-off location. The solution for $f$ and $v$ beyond $N_{tip}$ is set to zero. For subsequent time steps, $N_{tip}$ has two possibilities -- it can either advance or recede, which requires the following two conditions. (\textit{ii}) \textit{Advancing jet:} The values of $f$ at nodes $N_{tip}-1$ and $N_{tip}$ are extrapolated to find $f$ at $N_{tip}+1$. If the extrapolated value is larger than a predefined value of $5 \cdot 10^{-3}$, the parameter $N_{tip}$ is incremented by 1, and $f$ and $v$ at the new $N_{tip}$ are assigned values extrapolated from its previous two neighbours. (\textit{iii}) \textit{Receding jet:} If the value of $f$ at $N_{tip}$ falls below a predefined value of $10^{-3}$, $f$ and $v$ at $N_{tip}$ are set to zero and the parameter $N_{tip}$ is reduced by 1. These three conditions enable the numerical integration of the governing equations in a way that captures accurately the break-up of the jet and the motion of the tip.

A validation of the code is presented in Appendix \ref{app_1}. In the next section, we discuss the parameter values and domain size that we selected for our numerical simulations, in order to resemble the experimental conditions of Gumennik \textit{et al.}~\cite{gumennik2013}. 
\subsection{Numerical domain and parameter values} \label{NumModel}
\begin{figure}
\centering
   \includegraphics[width=\textwidth]{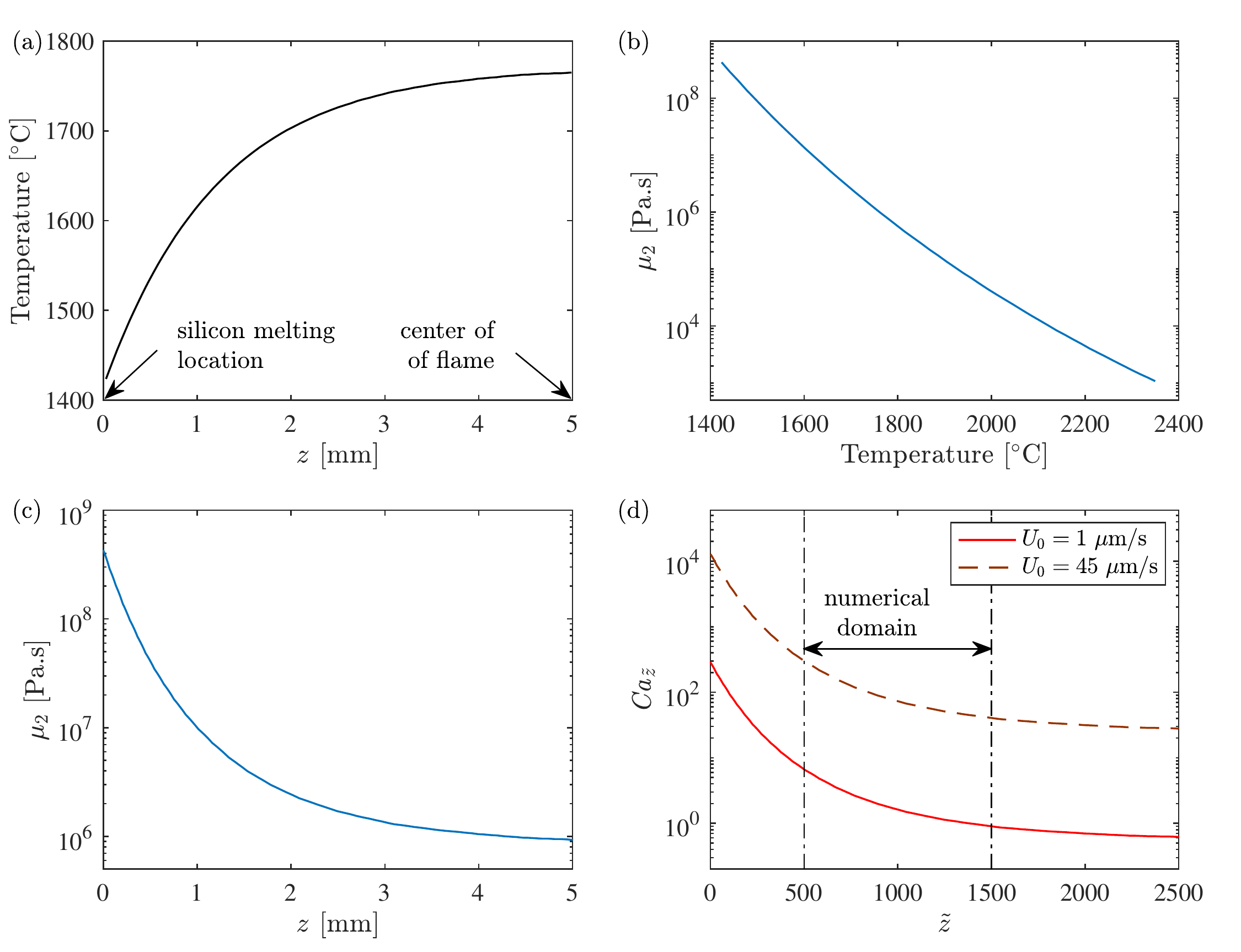}
 \caption{(a) Axial temperature profile between the liquefaction point of silicon and the center of the flame. (b) Silica viscosity $\mu_2$ as a function of temperature. Both plots are taken from Gumennik \textit{et al.}~\cite{gumennik2013}. (c) Silica viscosity profile along the axial direction. (d) Capillary number $\mathit{Ca}_{\tilde{z}}$ as a function of dimensionless axial coordinate, for $U_0 = 1 \, \mu$m/s and $45 \, \mu$m/s. The dash-dotted lines indicate the extent of the numerical domain.}
\label{figDataGum}
\end{figure}

We first deduce the values of $\mathit{We}$ and $\mathit{Ca}_{\tilde{z}}$ corresponding to the operating conditions of Gumennik \textit{et al.}~\cite{gumennik2013}. In their experiments, a silicon-in-silica co-axial fiber is fed into a flame at a constant  speed $U_0$, which varies between $1$ and $100 \, \mu$m/s. Since the flame is located slightly downstream of the inlet, the temperature of the co-axial fiber changes along its axial direction. Gumennik \textit{et al.}~\cite{gumennik2013} state that the temperature increases over a length of $5 \,$mm, from $T = 1414^\circ$C at the inlet, corresponding to the liquefaction point of silicon, to $T \simeq 1760^\circ$C in the heart of the flame. This affects the temperature-dependent silica viscosity $\mu_2$, which becomes a function of the axial coordinate. Correlating the temperature profile along the axial direction, shown in Figure \ref{figDataGum}(a), with the relationship between silica viscosity and temperature, shown in Figure \ref{figDataGum}(b), the profile of silica viscosity along the axial direction can be obtained in Figure \ref{figDataGum}(c). 
Note that the data in Figures \ref{figDataGum}(a)--(b) is from Gumennik \textit{et al.}~\cite{gumennik2013}. Observe that $\mu_2$ decreases by more than two orders of magnitude, from $10^8$ to $10^6$ Pa.s, over a few millimeters.

Based on the physical parameters, the Weber number $\mathit{We}$ lies between $10^{-11}$ and $10^{-13}$ depending on the feed speed $U_0$, which is computationally out of reach. Nevertheless, we show in Appendix \ref{app_2} that the break-up location and period are $\textit{We}$-independent in the numerically-tractable range $0.005<\textit{We}<0.1$. Thus, below a certain limit, the Weber number can be seen as a numerical artefact which has a negligible influence on the droplet size in comparison to the capillary number. We henceforth pick $\textit{We} = 0.05$ in our simulations, regardless of the feed speed $U_0$. The capillary number $\mathit{Ca}_{\tilde{z}}$ inherits the axial dependency of the silica viscosity $\mu_2(z)$, and therefore decreases by more than two orders of magnitude along the fiber. Furthermore, $\mathit{Ca}_{\tilde{z}}$ scales linearly with the feed speed $U_0$. For instance, as shown in Figure \ref{figDataGum}(d), $\mathit{Ca}_{\tilde{z}}$ decreases from $284$ to $0.62$ for $U_0=1 \, \mu$m/s, while it decreases from $12800$ to $27.9$ for $U_0=45 \, \mu$m/s. 

Finally, we restrict the size of the numerical domain considered in the simulations in order to render the computational time tractable, as explained in Appendix \ref{app_3}. Starting from the domain $\tilde{z} \in [0,2500]$ between the melting location of the silicon and the heart of the flame, we eliminate the region $\tilde{z} < 500$ in order to avoid high capillary numbers $\mathit{Ca}_{\tilde{z}}$ that would require prohibitively expensive computations. We also ignore the region $\tilde{z} > 1500$ since the jet breaks up before then. This leads us to the truncated domain $\tilde{z} \in [500,1500]$ pictured in Figure \ref{figDataGum}(d), in which we perform all the simulations shown in the next section using $\mathit{We} = 0.05$ and $U_0$-dependent $\mathit{Ca}_{\tilde{z}}$ profiles, such as those overlaid in the same figure.
\subsection{Numerical results}
\begin{figure}
\centering
   \includegraphics[width=\textwidth]{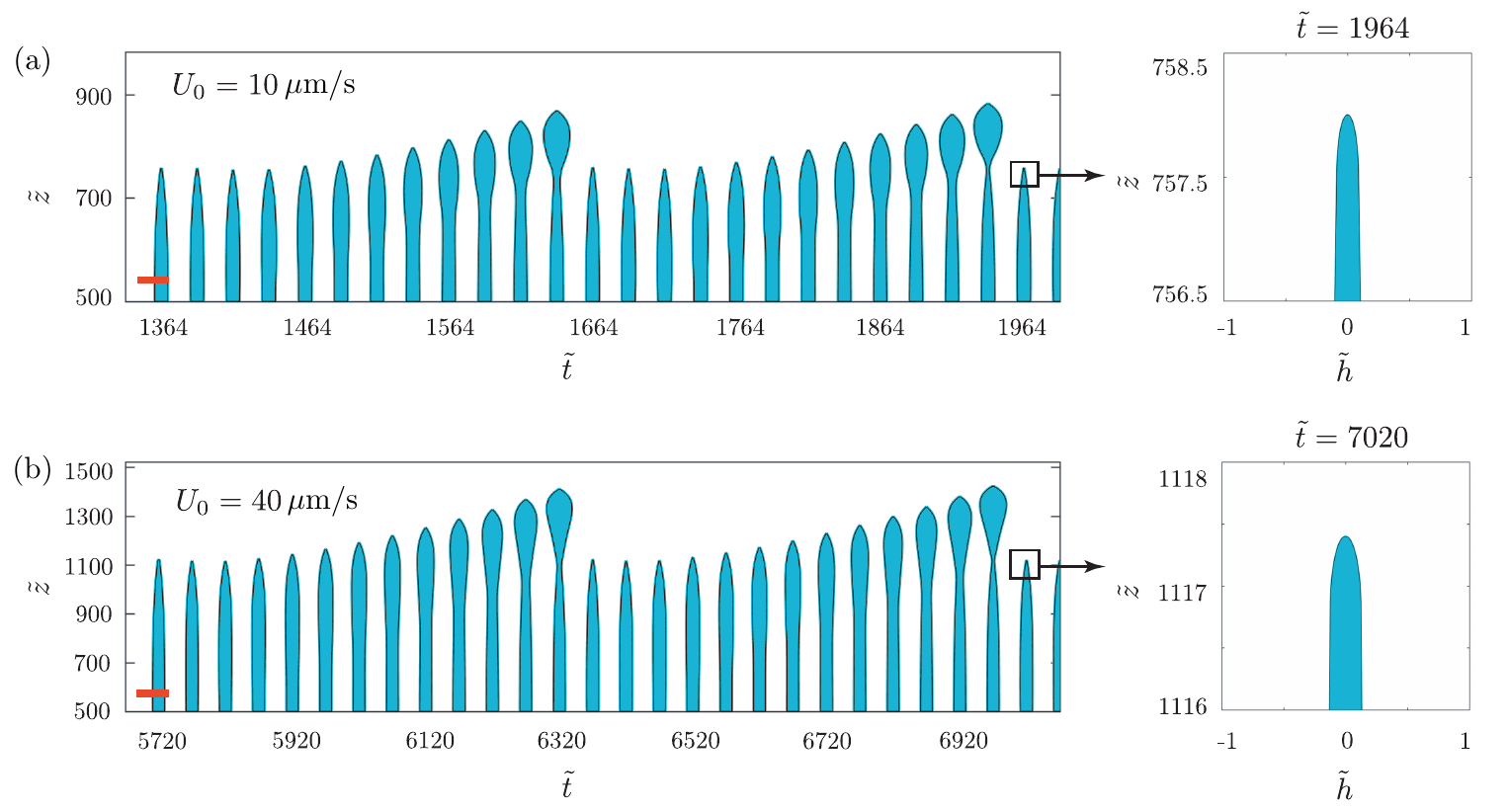}
\caption{Cascade plots of the evolution of the silicon-silica interface for feed speeds (a) $U_0 = 10 \, \mu$m/s and (b) $U_0 = 40 \, \mu$m/s. The dimensionless silicon core radius $\tilde{h} = h/h_0$ is plotted at fixed time intervals, and the red bar corresponds to a horizontal length scale of 5 dimensionless units. The magnified plots to the right show the shape of the jet tip right after break-up, with equal length scale for the horizontal and vertical axes.}
\label{Cascade}
\end{figure}

Using the numerical scheme described in Section \ref{NumScheme}, together with the numerical domain and parameter values presented in Section \ref{NumModel}, we compute solutions to the nonlinear governing equations \eqref{eqN2} for different feed speeds $U_0$. The simulations are run for a sufficiently long time to enter a quasi-steady regime wherein the jet breaks up at regular intervals of time and at the same axial location. In this regime, Figure \ref{Cascade} shows cascade plots of the evolution of the silicon-silica interface at fixed time intervals and over two consecutive break-up periods, for two different feed speeds of (a) 10$ \, \mu$m/s and (b) 40$ \, \mu$m/s. 
The interface is plotted in terms of the dimensionless silicon core radius $\tilde{h} = h/h_0$, and the red bar corresponds to a horizontal length scale of 5 dimensionless units. Note that the jets have very slender profiles -- their dimensionless inlet diameter is equal to $2$, whereas they travel over an axial distance of approximately one thousand. Indeed, for $U_0=10 \, \mu$m/s, the tip reaches a maximum dimensionless axial distance of $880$ and the break-up occurs at around $750$. With a higher feed speed of $40 \, \mu$m/s, the tip is capable of reaching a distance of 1400 with break-up taking place at around $1100$. The magnified plots to the right show the shape of the jet tip right after break-up, with equal length scale employed for the horizontal and vertical axes. Corresponding movies showing the jet break-up dynamics for these two feed speeds are included in the Supplementary Materials.

\begin{figure}
\centering
   \includegraphics[width=\textwidth]{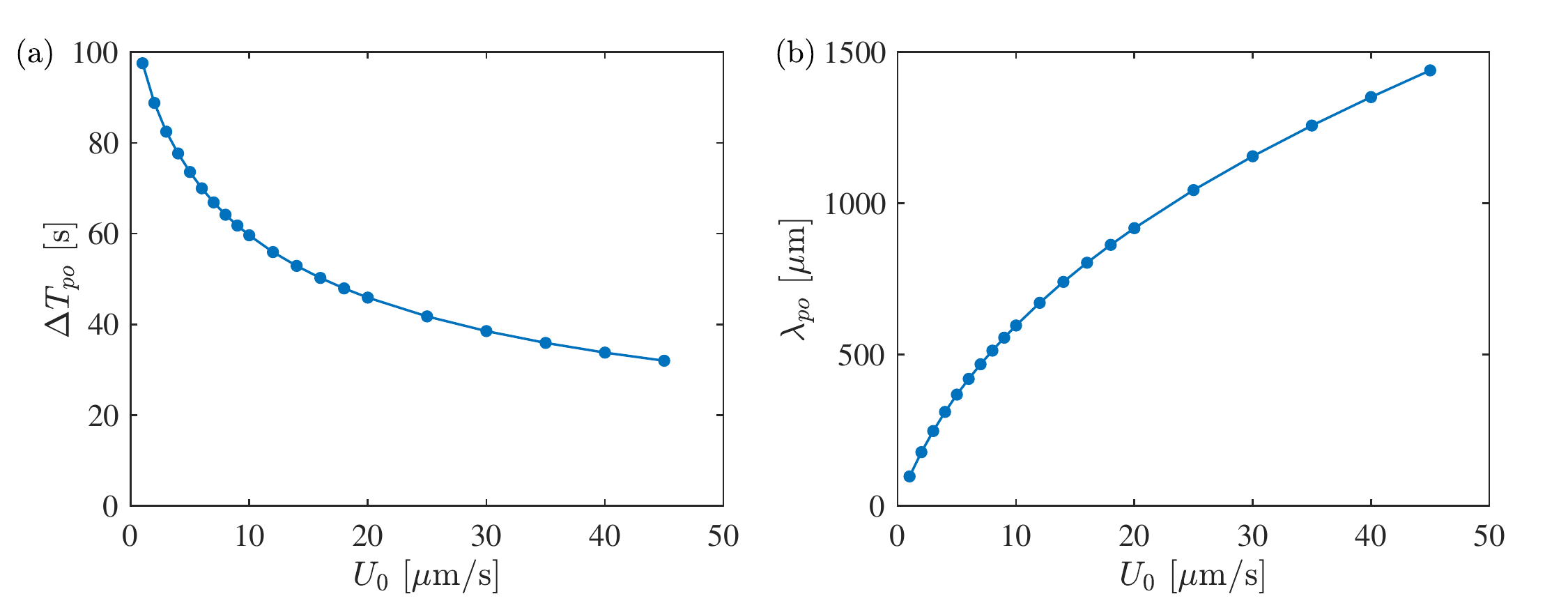}
\caption{(a) Break-up period $\Delta T_{po}$ and (b) distance travelled by the co-axial fiber over one break-up period, $\lambda_{po} = U_0 \Delta T_{po}$, as a function of the feed velocity $U_0$.}
\label{figAllData}
\end{figure}

In the quasi-steady regime, a minimum of eight consecutive break-up (or pinch-off) times $T_{po}$ are saved. These values are then used to calculate the break-up period $\Delta T_{po}$, which is defined as the average time between two consecutive pinch-offs. Figure \ref{figAllData} shows the dimensional break-up period $\Delta T_{po}$ as a function of the feed speed $U_0$, as well as the dimensional distance $\lambda_{po} = U_0 \Delta T_{po}$ traveled by the fiber over one break-up period.
In order to explain the decrease of $\Delta T_{po}$ with $U_0$, we recall from Figure \ref{Cascade} that as $U_0$ increases, the co-axial fiber travels farther into the domain and closer to the center of the flame. There, the lower silica viscosity results in enhanced capillary instability of the interface, causing faster jet break-up and hence smaller break-up periods as reported in Figure \ref{figAllData}(a). The sublinear trend displayed by the distance traveled $\lambda_{po}$ in Figure \ref{figAllData}(b) is also explained by the decrease of $\Delta T_{po}$ with $U_0$. Note that even though $\lambda_{po}$ reaches dimensions comparable to the size of the numerical domain, the break-up always occurs within the latter. This is because part of the mass influx between two consecutive break-ups contributes to a radial expansion of the silicon core, as seen in Figure \ref{Cascade}.

\begin{figure}
\centering
   \includegraphics[width=\textwidth]{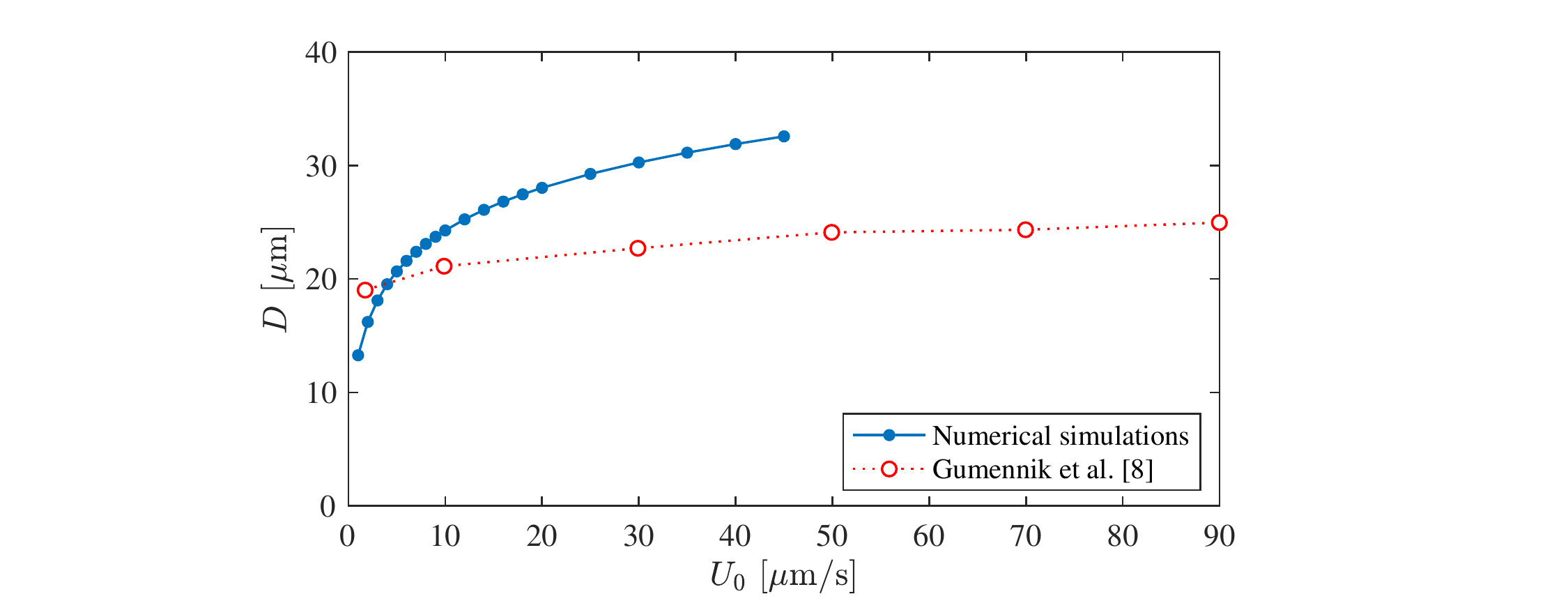}
\caption{Mean silicon sphere diameter $D$ as a function of the feed speed $U_0$. Comparison between our numerical simulations of \eqref{eqN2} and experimental data by Gumennik \textit{et al.}~\cite{gumennik2013}.}
\label{figComp}
\end{figure}

In order to compare our numerical observations with the experimental results of Gumennik \textit{et al.}~\cite{gumennik2013}, we calculate the diameter $D$ of the silicon spheres resulting from the break-up process using the mass conservation equation
\begin{equation}\label{eqN4}
\pi h_0^2 \lambda_{po}= \frac{\pi}{6} D^3.
\end{equation}
Figure \ref{figComp} displays the sphere diameter $D$ as a function of the feed speed $U_0$ for our simulations and for the experiments of Gumennik \textit{et al.}~\cite{gumennik2013}. We observe a good qualitative agreement between the two sets of data, with the governing equations \eqref{eqN2} being able to capture the increase in sphere diameter with feed speed, as well as its saturation at high feed speeds. Furthermore, the drop diameter predicted by the numerics is roughly comparable in magnitude to that observed in experiments, which is remarkable given that not a single fitting parameter has been used in our calculations.

There are different reasons that could explain the discrepancy between our numerical results and the experiments. First, although the temperature profile that we considered in Figure \ref{figDataGum}(a) comes from Gumennik \textit{et al.}~\cite{gumennik2013}, it was not directly measured from their experiments. Second, we noted in Appendix \ref{app_3} that truncating part of the entrance region out of the numerical domain results in a significant -- albeit unavoidable -- error at low feed speeds. Yet, this error becomes negligible for larger feed speeds, and hence we mostly attribute the discrepancy between the results to the effects of surface tension. The latter is assumed to be equal to $1.5 \,$N/m; however, Gumennik \textit{et al.}~\cite{gumennik2013} evaluated this value based on Ref.~\cite{kroll2006nano}, in which a range $\gamma=1.5 \pm0.3 \,$N/m is actually given. In fact, we show in Appendix \ref{app_4} that surface tension has a non-negligible effect on the sphere size, with the two being inversely proportional to each other. Additionally, we assumed that the surface tension at the silicon-silica interface remains constant over the entire temperature range of $1414-1760^{\circ}$C, unlike the viscosity of silica. In reality, studies show that the surface tension of silica in air \citep{kingery1959surface} and silicon in air \citep{shishkin2004surface,yuan2002surface,hibiya1998interfacial} can vary between $0.28-0.3 \,$N/m and $0.7-0.9 \,$N/m, respectively, over a temperature range of $1400-1800^{\circ}$C. Thus, a precise estimation of the surface tension at the silica-silicon interface could possibly lead to more accurate sphere size predictions.

\section{Conclusions and perspectives}
\label{sec:Conclusions}

In this article, we have tried to elucidate the physical mechanisms responsible for selecting the size of spherical silicon particles in the experimental setup of Gumennik \textit{et al.}~\cite{gumennik2013}. Such particles are obtained by feeding a silicon-in-silica co-axial fiber into a flame at a certain speed, triggering local melting of the silicon and Rayleigh-Plateau instability of the silicon-silica interface. We first derived a reduced model for the motion of the interface, consisting of two coupled one-dimensional nonlinear equations \eqref{eq:GoverningEquations}. Then, we analyzed the dynamics and dominant length scale of the instability that arises in this model using local linear stability analysis in its temporal and spatio-temporal flavors. Ultimately, however, we reached the conclusion that such linearized tools fail at predicting the particle size observed experimentally. Finally, we performed numerical simulations of the reduced nonlinear model. Without any adjustable parameters, we were able to recover in these simulations the particle size observed experimentally by Gumennik \textit{et al.}~\cite{gumennik2013}, as well as its qualitative behavior as the feed speed of the fiber is changed. 

Recalling the failure of the linear stability predictions, the success of the nonlinear analysis suggests that nonlinear effects play a predominant role in selecting the size of the silicon spheres. In other words, the break-up wavelength is largely independent of the initial growth of infinitesimal perturbations to the silicon-silica interface, which is contrary to the behavior of most pattern-forming systems \citep{cross1993,gallaire2017}. One might argue that the strong non-uniformity of the system -- imparted by the axial variation of silica viscosity over more than two orders of magnitude -- may explain the failure of local linear stability analysis. A global stability analysis would take such non-uniformity into account; however, numerical convergence of the resulting eigenvalue problem will be problematic due to the extreme variation in silica viscosity. Nonetheless, we are confident that the good agreement observed between the numerical simulations and the experiments in Figure \ref{figComp} is by and large attributable to the nonlinearity of the viscous term originating from the outer silica, as opposed to the axial variation of the silica viscosity itself. 

To prove this point, we compared in Appendix \ref{app_5} numerical simulations of equations \eqref{eqN2} for a silicon-in-silica fiber at constant capillary number with numerical simulations of equations \eqref{appEq1} for a jet in an inert medium. Both simulations were performed in the low Weber number limit $\mathit{We} = 0.01$, and using $\mathit{Ca} = 1$ and $1/3$ for \eqref{eqN2} and \eqref{appEq1}, respectively. In this way, the dispersion relations of both equations are identical -- that is, their linear stability properties are indistinguishable. Even so, we were surprised to observe that their nonlinear behaviors are markedly different: as shown in Figure \ref{HolViscComp}, equations \eqref{eqN2} for the silicon-in-silica fiber produce regularly-spaced droplets, while equations \eqref{appEq1} for the jet in an inert medium lead to the formation of one ever-growing pendant drop. Given that the only difference between these two sets of equations is the nonlinear form of the viscous term, we conclude that the length scale of the droplets produced in the silicon-in-silica fiber is really set by the nonlinearity of the viscous contribution from the outer silica \footnote{Although a nonlinear viscous term might sound paradoxical due to the linearity of viscous diffusion, it is worth remembering that here, it is the geometric nonlinearity of the silicon-silica interface that makes the outer velocity field -- hence the viscous diffusion -- a nonlinear function of the interface position.}. Thus, we hypothesize that the latter might amount to some kind of body force that pinches off droplets once they grow big enough, in the same spirit as the dynamics of a dripping faucet \citep{michael1976,peregrine1990}.

\appendix

\section{Validity of constant outer pressure assumption} \label{app_0}

\begin{figure}
\centering
\includegraphics[width=\textwidth]{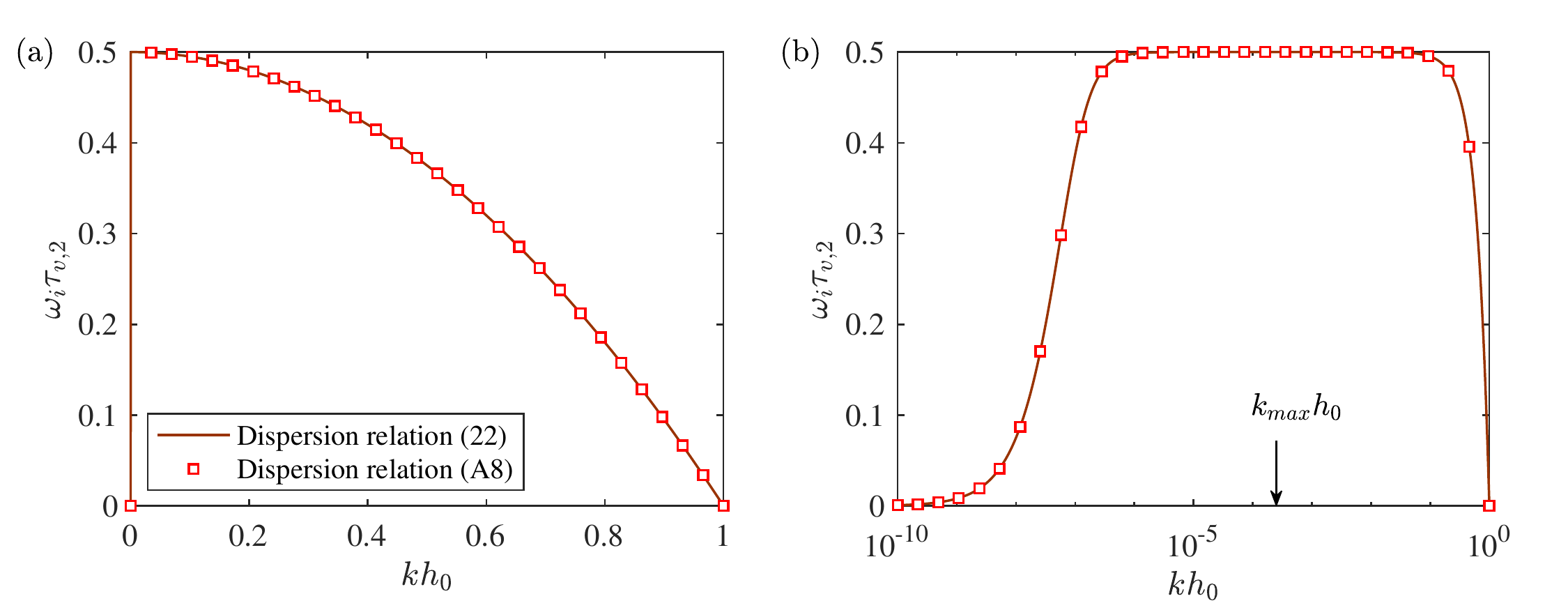}
\caption{Temporal growth rate $\omega_i$ as a function of the real wavenumber $k$ from the dispersion relations \eqref{eq:DispersionRelation} and \eqref{eq:FullDispersionRelation}, in (a) linear and (b) logarithmic wavenumber scale. The growth rates of both \eqref{eq:DispersionRelation} and \eqref{eq:FullDispersionRelation} are maximum at $k_{max} h_0 \simeq 2.40 \cdot 10^{-4}$.}
\label{fig:TemporalDispersionRelationAppendix}
\end{figure}

In this appendix, we show that solving explicitly for the outer silica pressure -- instead of assuming that it is constant, as in Section \ref{sec:OuterSilicaCladding} -- leads to governing equations with a dispersion relation that is numerically identical with \eqref{eq:DispersionRelation}. First, we note that under the assumption that $u_2 = U_0$ and keeping only the terms with a linear contribution in the perturbation, the Navier-Stokes momentum equation \eqref{eq:NSOuter1} reduces to
\begin{equation}
\frac{\partial v_2}{\partial t} + U_0 \frac{\partial v_2}{\partial z} = -\frac{1}{\rho_2} \frac{\partial p_2}{\partial r} + \nu_2 \left( \frac{\partial^2 v_2}{\partial r^2} + \frac{\partial^2 v_2}{\partial z^2} + \frac{1}{r} \frac{\partial v_2}{\partial r} - \frac{v_2}{r^2} \right).
\label{eq:ReducedNSOuter}
\end{equation}
Next, we insert expression \eqref{eq:OuterVelocity} for $v_2$ inside \eqref{eq:ReducedNSOuter} and, like before, we only retain the terms with a linear contribution in the perturbation to get
\begin{equation}
\frac{h}{r} \left( \frac{\partial^2 h}{\partial t^2} + 2U_0 \frac{\partial^2 h}{\partial t \partial z} + U_0^2 \frac{\partial^2 h}{\partial z^2} \right) = -\frac{1}{\rho_2} \frac{\partial p_2}{\partial r} + \nu_2 \frac{h}{r} \frac{\partial^2}{\partial z^2} \left( \frac{\partial h}{\partial t} + U_0 \frac{\partial h}{\partial z} \right).
\label{eq:ReducedNSOuter2}
\end{equation}
(Here, we would like to point out that the viscous term does not cancel entirely, despite what is stated in section 3.5.1 of the review by Eggers and Villermaux \cite{eggers2008}. This is due to the axial dependency of the radial velocity field $v_2(r,z,t)$, inherited from the interface height $h(z,t)$ and overlooked by the aforementioned authors.) Equation \eqref{eq:ReducedNSOuter2} can now be integrated along $r$ to find an expression for the pressure $p_2$, provided one has a suitable boundary condition. The harmonicity of the pressure field ensures that radial and axial length scales are comparable, which implies that $p_2$ decays exponentially in the radial direction over a length scale $\lambda \sim 1/k$ when the interface is deformed by a wavenumber $k$. Since we are ultimately looking for the dispersion relation of the system, we thus consider that the pressure vanishes at $r = h + 1/k$ and integrate \eqref{eq:ReducedNSOuter2} to obtain
\begin{equation}
p_2 = \rho_2 h \left( \frac{\partial^2 h}{\partial t^2} + 2U_0 \frac{\partial^2 h}{\partial t \partial z} + U_0^2 \frac{\partial^2 h}{\partial z^2} \right) \ln \left( \frac{h}{r} + \frac{1}{kr} \right) - \mu_2 h \frac{\partial^2}{\partial z^2} \left( \frac{\partial h}{\partial t} + U_0 \frac{\partial h}{\partial z} \right) \ln \left( \frac{h}{r} + \frac{1}{kr} \right).
\label{eq:SilicaPressure}
\end{equation}
Finally, we plug the above expression for $p_2$ into the normal stress condition \eqref{eq:ReducedNormalStressBC}, which yields an expression for the leading-order inner pressure,
\begin{align}
\bar{p}_{10} &= \underbrace{\rho_2 h \left( \frac{\partial^2 h}{\partial t^2} + 2U_0 \frac{\partial^2 h}{\partial t \partial z} + U_0^2 \frac{\partial^2 h}{\partial z^2} \right) \ln \left( 1 + \frac{1}{kh} \right)}_{\substack{\text{inertial term from} \\ \text{pressure in outer silica}}} - \underbrace{\mu_2 h \frac{\partial^2}{\partial z^2} \left( \frac{\partial h}{\partial t} + U_0 \frac{\partial h}{\partial z} \right) \ln \left(1 + \frac{1}{kh} \right)}_{\substack{\text{viscous term from} \\ \text{pressure in outer silica}}} \nonumber \\
&\quad + \underbrace{\gamma \kappa}_{\substack{\text{Laplace} \\ \text{pressure jump}}} + \underbrace{\frac{2 \mu_2}{h} \left( \frac{\partial h}{\partial t} + U_0 \frac{\partial h}{\partial z} \right)}_{\substack{\text{normal component of} \\ \text{viscous stress in silica}}}.
\label{eq:FullPressure}
\end{align}
Compared with the expression \eqref{eq:InnerPressure} we obtained earlier, there are here two additional contributions to the inner pressure $\bar{p}_{10}$. Recalling that the radial length scale $h_0$ of the jet is much smaller than its axial length scale $\lambda \sim 1/k$, we have $kh_0 \ll 1$ and a dominant balance comparison between the two viscous contributions gives
\begin{equation}
\frac{\text{viscous term from pressure in silica}}{\text{normal component of viscous stress in silica}} \sim \frac{1}{2} \ln \left( 1+\frac{1}{kh_0} \right) (kh_0)^2 \ll 1.
\end{equation}
The viscous term inherited from the silica pressure $p_2$ can therefore be neglected in \eqref{eq:FullPressure}, leading to the simplified expression
\begin{align}
\bar{p}_{10} &= \rho_2 h \left( \frac{\partial^2 h}{\partial t^2} + 2U_0 \frac{\partial^2 h}{\partial t \partial z} + U_0^2 \frac{\partial^2 h}{\partial z^2} \right) \ln \left( 1 + \frac{1}{kh} \right) + \gamma \kappa + \frac{2 \mu_2}{h} \left( \frac{\partial h}{\partial t} + U_0 \frac{\partial h}{\partial z} \right).
\label{eq:Pressure}
\end{align}
Combining the above expression with \eqref{eq:OneDimensionalEquation1} and \eqref{eq:OneDimensionalEquation2} yields
\begin{subequations}
\begin{align}
\frac{\partial \bar{u}_{10}}{\partial t} + \bar{u}_{10} \frac{\partial \bar{u}_{10}}{\partial z} &= \rho_2 h \left( \frac{\partial^2 h}{\partial t^2} + 2U_0 \frac{\partial^2 h}{\partial t \partial z} + U_0^2 \frac{\partial^2 h}{\partial z^2} \right) \ln \left( 1 + \frac{1}{kh} \right) \nonumber \\
&\quad - \frac{\gamma}{\rho_1} \frac{\partial \kappa}{\partial z} - \frac{2}{\rho_1} \frac{\partial}{\partial z} \left[ \frac{\mu_2}{h} \left( \frac{\partial h}{\partial t} + U_0 \frac{\partial h}{\partial z} \right) \right], \\
\frac{\partial h}{\partial t} + \bar{u}_{10} \frac{\partial h}{\partial z} &= - \frac{1}{2} \frac{\partial \bar{u}_{10}}{\partial z} h,
\end{align}
\label{eq:FullGoverningEquations}%
\end{subequations}%
where $\kappa$ is given by \eqref{eq:Curvature2}, and the log term is new compared with the governing equations \eqref{eq:GoverningEquations} obtained earlier by neglecting $p_2$. Note that due to the presence of the wavenumber $k$ in the log term, equations \eqref{eq:FullGoverningEquations} are not governing equations in the true sense. Nonetheless, the corresponding dispersion relation is
\begin{align}
&\frac{\rho_1 h_0^3}{\gamma} \left[ 1 + \frac{1}{2} \frac{\rho_2}{\rho_1} \ln \left( 1 + \frac{1}{kh_0} \right) (kh_0)^2 \right] (\omega - U_0 k)^2 \nonumber \\
&\quad+ i \frac{\mu_2 h_0}{\gamma} (kh_0)^2 (\omega - U_0 k) + \frac{1}{2} \left[ (kh_0)^2-(kh_0)^4 \right] = 0.
\label{eq:FullDispersionRelation}
\end{align}
In Figure \ref{fig:TemporalDispersionRelationAppendix}, we plot the dispersion relations \eqref{eq:DispersionRelation} and \eqref{eq:FullDispersionRelation} under the same conditions as in Section \ref{sec:LocalTemporalStability}. The two dispersion relations are virtually indistinguishable from each other, and the maximum growth rate happens at $k_{max} h_0 \simeq 2.40 \cdot 10^{-4}$ in both cases. This validates our assumption that $p_2$ is approximately constant -- as far as linearized dynamics are concerned, at least. Indeed, the additional log term appearing in \eqref{eq:FullGoverningEquations} is nonlinear and could possibly affect the nonlinear behavior of the system. Ultimately, though, the good agreement we have obtained with the results of Gumennik \textit{et al.}~\cite{gumennik2013} in Figure \ref{figComp} lends confidence to the constant outer pressure assumption.

\section{Numerical code validation on a jet in an inert medium} \label{app_1}
We validate our numerical code with simulations of the reduced governing equations obtained by Eggers and Dupont \cite{eggers1994} for a jet with density $\rho$ and viscosity $\mu$ in an inert medium. These one-dimensional equations are obtained from the same long-wavelength approximation that we have used to derive the governing equations \eqref{eqN2} of the silicon-in-silica jet. Written in the same nondimensional variables $f = (h/h_0)^2$ and $v = \bar{u}_{10}/U_0$ as in Section \ref{sec:NonlinearStabilityAnalysis}, they take the form
\begin{subequations}
\begin{align}
\mathit{We} \left(\frac{\partial v}{\partial \tilde{t}} + v\frac{\partial v}{\partial \tilde{z}}\right ) &=  - \frac{\partial \tilde{\kappa}}{\partial \tilde{z}} +  \frac{3\mathit{Ca}}{f} \frac{\partial}{\partial \tilde{z}} \left( f \frac{\partial v}{\partial \tilde{z}} \right),
\label{subeqappEq1}\\
\frac{\partial f}{\partial \tilde{t}} &= -\frac{\partial (fv)}{\partial \tilde{z}},
\\ 
\tilde{\kappa} &= \frac{(2-f'')f + f'^2}{2 ( f'^2/4 + f)^{3/2}}.
\end{align}
\label{appEq1}%
\end{subequations}%
Here, $\mathit{We} = {\rho h_0 {U_0}^2}/{\gamma}$ and $\mathit{Ca} = {\mu U_0}/{\gamma}$, and $z$ and $t$ refer to the dimensionless axial coordinate and time, respectively. Observe that the only difference between these equations and equations \eqref{eqN2} for the silicon-in-silica jet consists in the exact expression of the nonlinear viscous term (that which contains the capillary number). In \eqref{appEq1}, the viscous term originates from the axial velocity of the jet, while in \eqref{eqN2} it is due to the radial velocity of the outer silica. Nevertheless, as pointed out in Section \ref{sec:DispersionRelation}, the linear dispersion relations associated with \eqref{appEq1} and \eqref{eqN2} are identical, save for a factor 3 multiplying $\mathit{Ca}$.

Direct numerical simulations of equations \eqref{appEq1} have been performed by van Hoeve \textit{et al.}~\cite{Hoeve2010} and validated against experiments. Their numerical results describe micro-jets of initial radius $h_0 = 18.5 \, \mu$m with density $\rho = 1098 \, \mathrm{kg}/\mathrm{m}^3$,  viscosity $\eta = 3.65 \,$mPa.s, and surface tension $\gamma = 67.9 \,$mN/m. The jet is injected at a constant flow rate $Q = 0.35 \,$mL/min, corresponding to an initial jet velocity $U_0 = Q/(\pi h_0^2) = 5.4 \,$m/s. The flow can thus be described by the dimensionless numbers $\mathit{Ca}=0.295$ and $\mathit{We}=8.7$. 
To initiate jet break-up in their numerical simulations, a harmonic modulation of the dimensional nozzle radius is applied as follows:
\begin{equation}
h(z=0,t) = h_0 + \delta \sin 2\pi nt,
\end{equation}
with $\delta/h_0 \approx 0.005$ the forcing amplitude, and $n$ the driving frequency. The latter is selected to match the optimum wavelength $\lambda_{opt}$ for jet breakup, that is, $n = U_0/\lambda_{opt}$. To ensure a constant flow rate $Q$ through the nozzle, the dimensional velocity is modulated correspondingly as
\begin{equation}
\bar{u}_{10}(z=0,t) = \frac{h_0^2 U_0}{[h(z=0,t)]^2}.
\end{equation}
The amplitude of the wave imparted by the forcing at the nozzle grows until it equals the radius of the jet. Pinch-off or jet break-up is then defined as when the minimum width of the jet is below a predefined value set to $10^{-3} h_0$.

\begin{figure}
	\centering
	\includegraphics[width=0.8\textwidth]{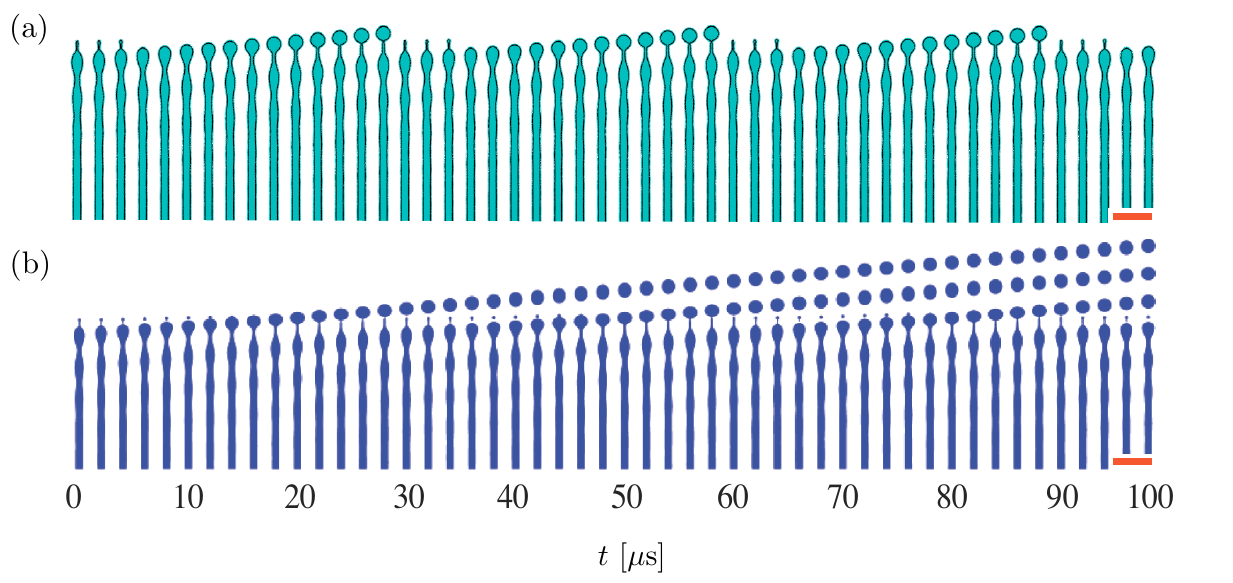}
	\caption{Comparison of numerical simulations of the governing equations \eqref{appEq1} for a jet in an inert medium with $\mathit{Ca}=0.295$ and $\mathit{We}=8.7$. Results from (a) our numerical scheme, described in Section \ref{NumericalScheme}, and (b) van Hoeve \textit{et al.}~\cite{Hoeve2010}. The jet interface is plotted every $2 \, \mu$s, and the red bar corresponds to a length scale of $200 \, \mu$m. }   
	\label{fighoeve}
\end{figure}

We proceed to the validation of the numerical scheme described in Section \ref{NumericalScheme}, by computing solutions to the governing equations \eqref{appEq1} with the same harmonic forcing and parameter values as in van Hoeve \textit{et al.}~\cite{Hoeve2010}. A hemispherical droplet described by $h=({h_0}^2 -z^2)^{1/2}$ is used as initial condition for the shape of the jet, the tip of which is therefore initially at $z=h_0$. The velocity is initialized to $U_0$ everywhere along the jet.  A fixed number of grid points, corresponding to a discretization size $d\tilde{z} = 0.05$, is uniformly distributed throughout the entire domain. The final validation is presented in Figure \ref{fighoeve}, which shows a time series of the dynamics of jet break-up based on (\textit{a}) our numerical scheme and (\textit{b}) the numerical results from van Hoeve \textit{et al.}~\cite{Hoeve2010}.
For both figures, the evolution of the jet shape is shown at time intervals of $2 \, \mu$s. Our numerical model predicts a break-up period of  $25 \, \mu$s and a break-up length of $ 856 \, \mu$m. The results of van Hoeve \textit{et al.}~\cite{Hoeve2010}, on the other hand, have a break-up period of about $26$ to $30 \, \mu$s and a break-up length of about $800 \, \mu$m.  The error in break-up length between the two codes can be explained by the difference in grid size. Overall, Figure \ref{fighoeve} shows a good agreement between both results and validates our numerical scheme and implementation.

\section{Silicon-in-silica fiber with constant capillary number} \label{app_2}
In this appendix, we perform numerical simulations of the governing equations \eqref{eqN2} for the silicon-in-silica co-axial fiber, but using a \textit{constant} capillary number $\mathit{Ca}$. Such an assumption serves as a basis for understanding the behavior of the real system with spatially-varying capillary number $\mathit{Ca}_{\tilde{z}}$. Specifically, our goal here is two-fold: we show that the Weber number is a numerical artefact provided $\mathit{We}$ is small enough, and we study the numerical convergence of our scheme.

We compute the jet break-up characteristics for different values of $\mathit{Ca} \in [0.1,2]$ and $\mathit{We} \in [0.005,0.1]$. The simulation time is kept sufficiently large (about $1000$ dimensionless time units) to obtain a quasi-steady regime where drops are formed at regular intervals of time and at the same distance from the nozzle exit. The domain size is fixed at $50h_0$ for low capillary numbers but is progressively increased for higher capillary numbers. Indeed, higher capillary numbers correspond to increased viscous effects, slowing down the growth of interface perturbations and resulting in droplets forming further away from the nozzle.

\begin{figure}
\centering
   \includegraphics[width=\textwidth]{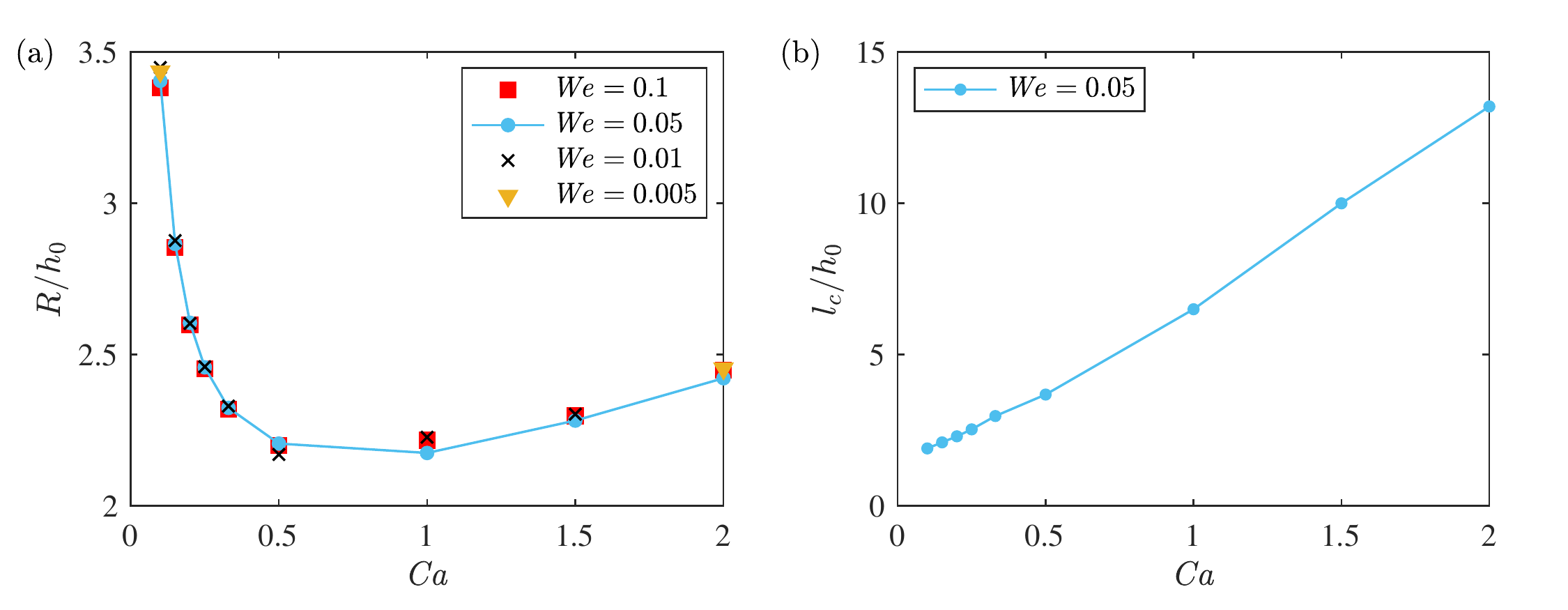}
\caption{(a) Dimensionless drop radius $R/h_0$ as a function of constant capillary number $\mathit{Ca}$, for different values of Weber number $\mathit{We} \in [0.005,0.1]$. (b) Dimensionless break-up length $l_c/h_0$ as a function of constant capillary number for $\mathit{We}=0.05$. The break-up dynamics resembles dripping at low $\mathit{Ca}$ numbers and jetting at higher $\mathit{Ca}$ numbers. }
\label{figCaEffect}
\end{figure}

Figure \ref{figCaEffect}(a) reports the break-up radius as a function of $\mathit{Ca}$, for different values of $\mathit{We}$.
Clearly, the break-up characteristics are $\mathit{We}$-independent for $\mathit{We} \le 0.05$. Thus, approximating the break-up characteristics for any $\mathit{We}$ smaller than $0.05$ with the corresponding values at $\mathit{We}=0.05$ is a valid assumption, which we extend in Section \ref{sec:NonlinearStabilityAnalysis} to the case of spatially-varying capillary number. 

Note, interestingly, that the break-up period follows a non-monotonous trend as $\mathit{Ca}$ is increased from $0.1$ to $2$. As shown in Figure \ref{figCaEffect}(b), the break-up occurs further away from the nozzle as $\mathit{Ca}$ is increased, in a way that is reminiscent of a transition from dripping to jetting \citep{utada2007}. It could therefore be possible that the non-monotonicity of the curve in Figure \ref{figCaEffect}(a) is related to an absolute to convective instability transition \citep{guillot2007}. More research is needed to confirm this assertion, however, and this goes beyond the scope of this paper.

\begin{figure}
\centering
   \includegraphics[width=\textwidth]{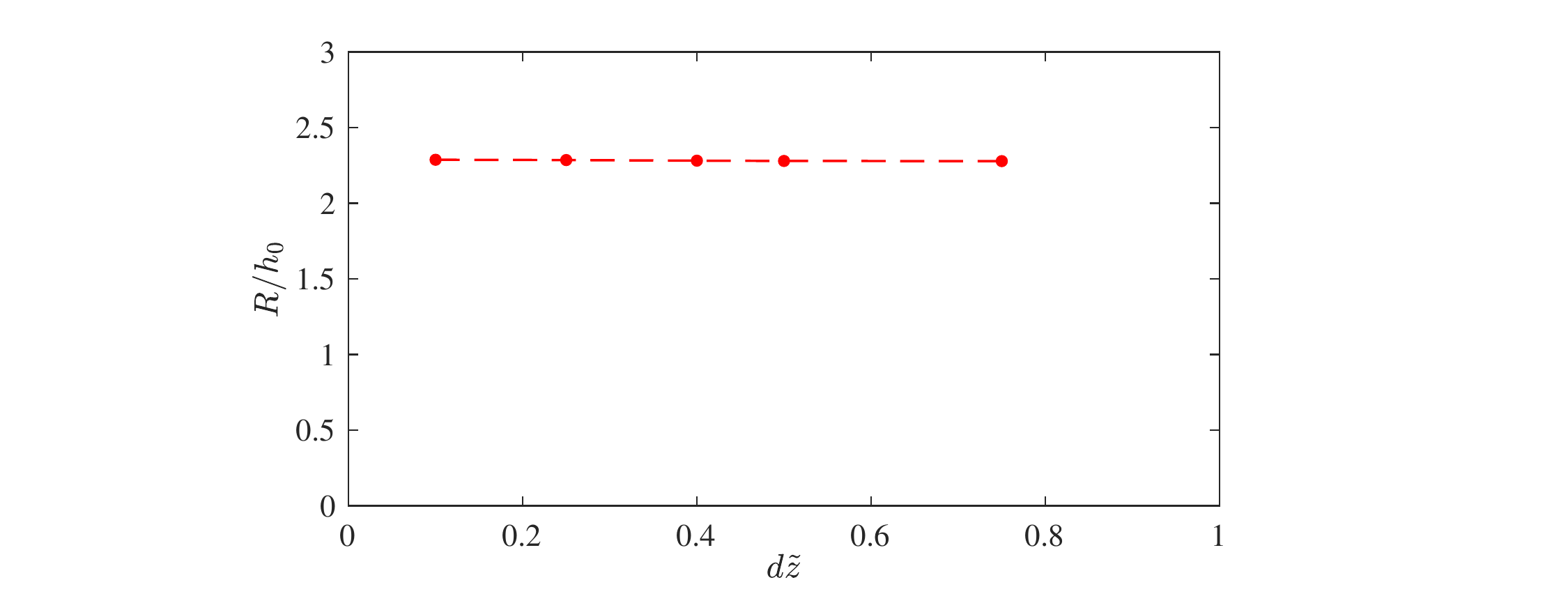}
\caption{Drop radius as a function of grid size for $\mathit{Ca}=1.5$ and $\mathit{We}=0.05$. The results show a weak dependence of the break-up characteristics on the grid size.}
\label{gridDepCstCa}
\end{figure}

Finally, a grid size-dependency test was performed for various values of $\mathit{Ca}$ and $\mathit{We} = 0.05$. It was observed that the break-up period and hence the drop radius have a weak dependence on the grid size, as shown in Figure \ref{gridDepCstCa} for the case $\mathit{Ca}=1.5$.
As the nondimensional grid size $d\tilde{z}$ is increased from $0.1$ to $0.75$, the drop radius $R$ decreases by merely 0.39\%. Thus, we selected grid sizes $d\tilde{z}$ comprised between $0.1$ and $0.56$ for the simulations presented in this appendix, and between $0.45$ and $0.65$ for the simulations in Section \ref{sec:NonlinearStabilityAnalysis}.

\section{Selection of a truncated numerical domain}\label{app_3}

In this appendix, we describe how we select a restricted region of the total physical domain for the numerical simulations in Section \ref{sec:NonlinearStabilityAnalysis}, in order to balance computational cost and accuracy. Experimental observations from Gumennik \textit{et al.}~\cite{gumennik2013} show that the jet always breaks up before reaching the heart of the flame. As a starting point, we thus restrict our attention to the $5$-mm-long region between the inlet and the heart of the flame, which we denote $\tilde{z} \in [0,2500]$. 

In addition, we have to alter the entrance location of the numerical domain, due to the fact that our numerical scheme can only work robustly with capillary numbers $\mathit{Ca}_{\tilde{z}}$ below $400$. As seen in Figure \ref{figDataGum}(d), for higher values of $U_0$ this limit is clearly exceeded at $\tilde{z} = 0$. Thus, with the aim of computing drop characteristics for feed speeds up to $U_0=50 \, \mu$m/s, we decide to reduce the domain size to $\tilde{z} \in [500,2500]$. In this way, the capillary number at $\tilde{z}=500$ for $U_0=50 \, \mu$m/s is 332, well within the computational limit. Eliminating the region $\tilde{z} \in [0,500]$ is a reasonable approximation since the silica viscosity in this region is large enough that the jet instability will not grow appreciably. Indeed, for feed speed $U_0=25 \, \mu$m/s, moving the entrance location from $\tilde{z} = 500$ to $400$ and $200$ produces a relative difference in sphere size of $4.4\%$ and $10\%$ while the corresponding computational cost increases 2-fold and 9-fold, respectively. 

\begin{figure}
\centering
   \includegraphics[width=\textwidth]{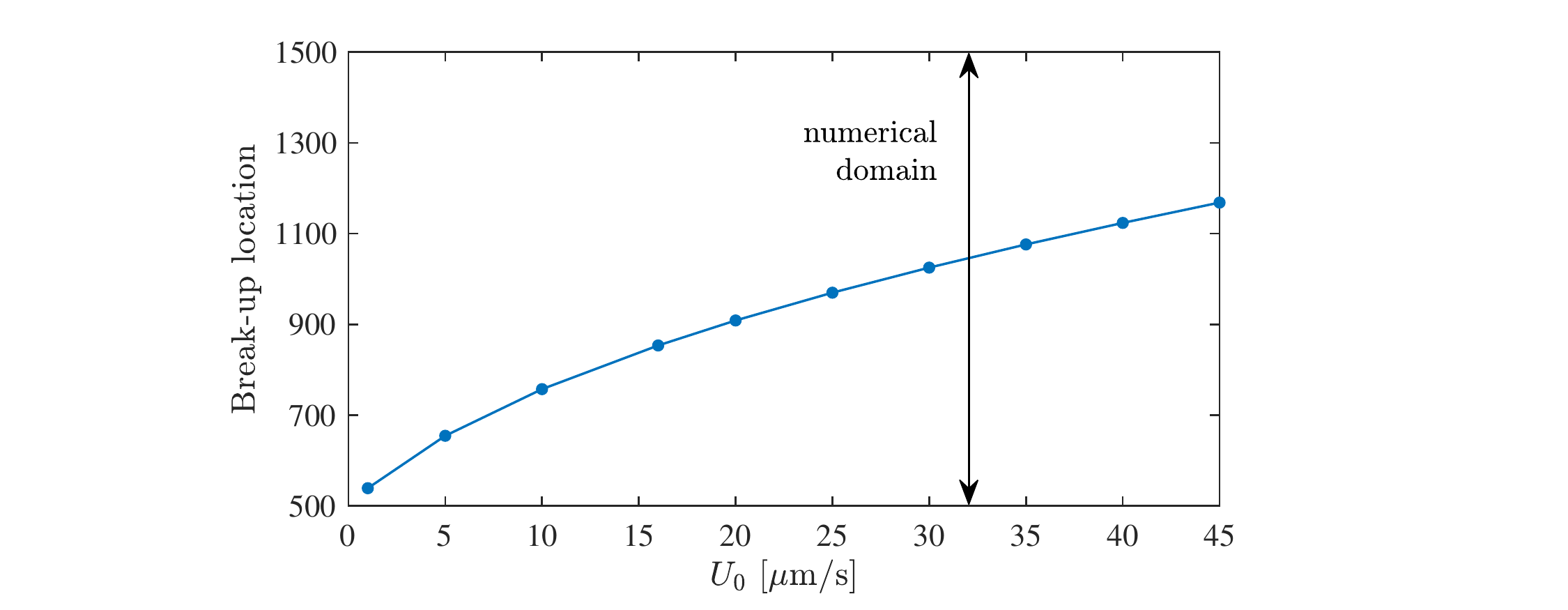}
\caption{Break-up location as a function of feed speed $U_0$ for the truncated numerical domain $\tilde{z} \in [500,1500]$. While the break-up always takes place within the truncated domain, its location progressively moves downstream and shifts towards the end of the domain as $U_0$ is increased.} \label{figLcz1000}
\end{figure}

Finally, we reduce the numerical domain size to $\tilde{z} \in [500,1500]$ on the assumption that the silica viscosity in the region $\tilde{z}<1500$ is sufficiently low to capture jet break-up. This assumption is verified by analyzing the break-up location as a function of the feed speed. Figure \ref{figLcz1000} shows that for feed speeds in the range of $1-45 \, \mu$m/s, the jet breaks up within the truncated numerical domain $\tilde{z} \in [500,1500]$. 
Furthermore, we verified that the sphere radius obtained with feed speeds $1$, $10$, and $40 \, \mu$m/s did not change between domains $\tilde{z} \in [500,1500]$ and $\tilde{z} \in [500,1700]$.

\section{Effect of surface tension on sphere size}\label{app_4}
\begin{figure}
\centering
   \includegraphics[width=\textwidth]{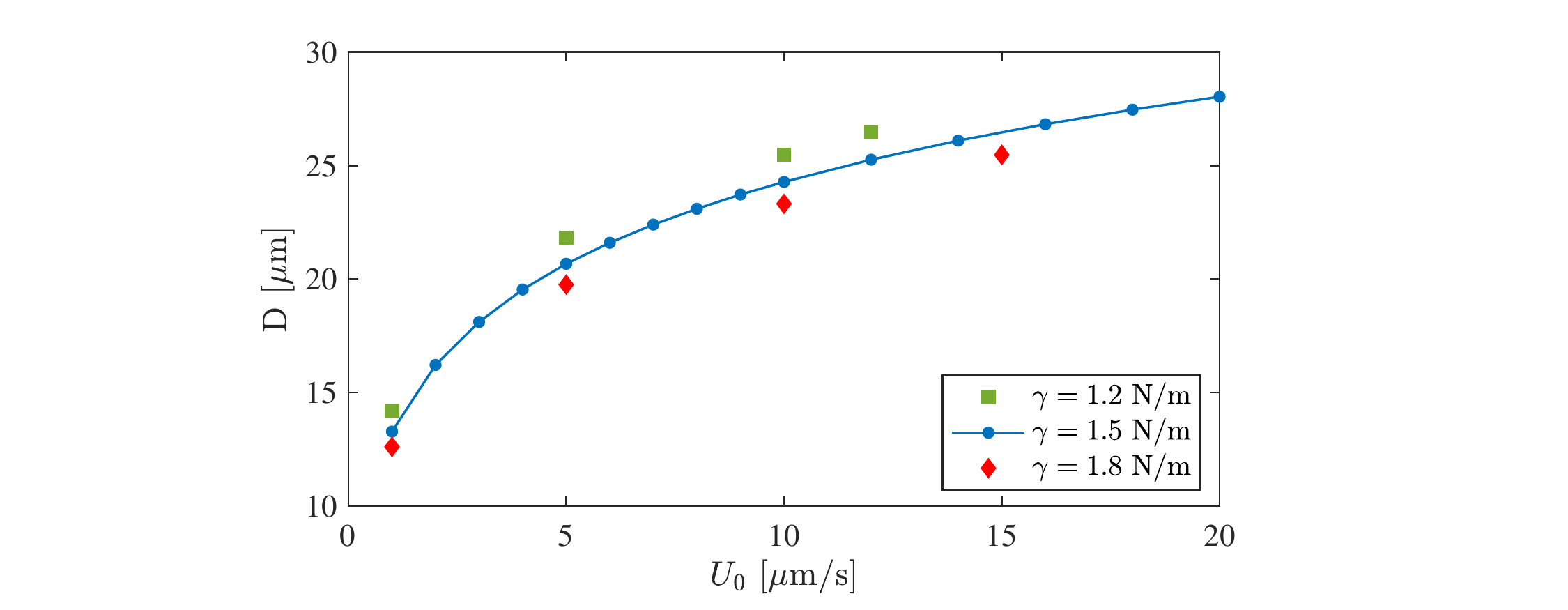}
\caption{Effect of surface tension on drop size. Relative errors of 20\% in the magnitude of the surface tension can cause corresponding errors of 4--5\% in the drop diameter.}
\label{fig:SurfaceTensionEffect}
\end{figure}

Here, we evaluate the effect of surface tension between silicon and silica on particle size, as predicted by our model. Figure \ref{fig:SurfaceTensionEffect} shows the predicted sphere diameter for three different values of the surface tension, all comprised within the error range provided by Ref.~\cite{kroll2006nano}. It is observed that the sphere size is inversely proportional to the surface tension. For example, for a feed speed of $10 \, \mu$m/s, decreasing the surface tension by 20\% from $\gamma=1.5$ N/m to $1.2$ N/m increases the predicted sphere diameter by 4.9\%, from $D = 24.3 \, \mu$m to $25.5 \, \mu$m. Conversely increasing the surface tension by 20\% from $\gamma=1.5$ N/m to $1.8$ N/m decreases the predicted sphere diameter by 4.1\%, from $D = 24.3 \, \mu$m to $23.3 \, \mu$m. This is not surprising since a higher value of surface tension implies more driving force for the pinching, hence faster break-ups that result into smaller drop sizes. 

\section{Comparison of nonlinear behaviors of silicon-in-silica fiber at constant capillary number and viscous jet}\label{app_5}

\begin{figure}
\centering
   \includegraphics[width=\textwidth]{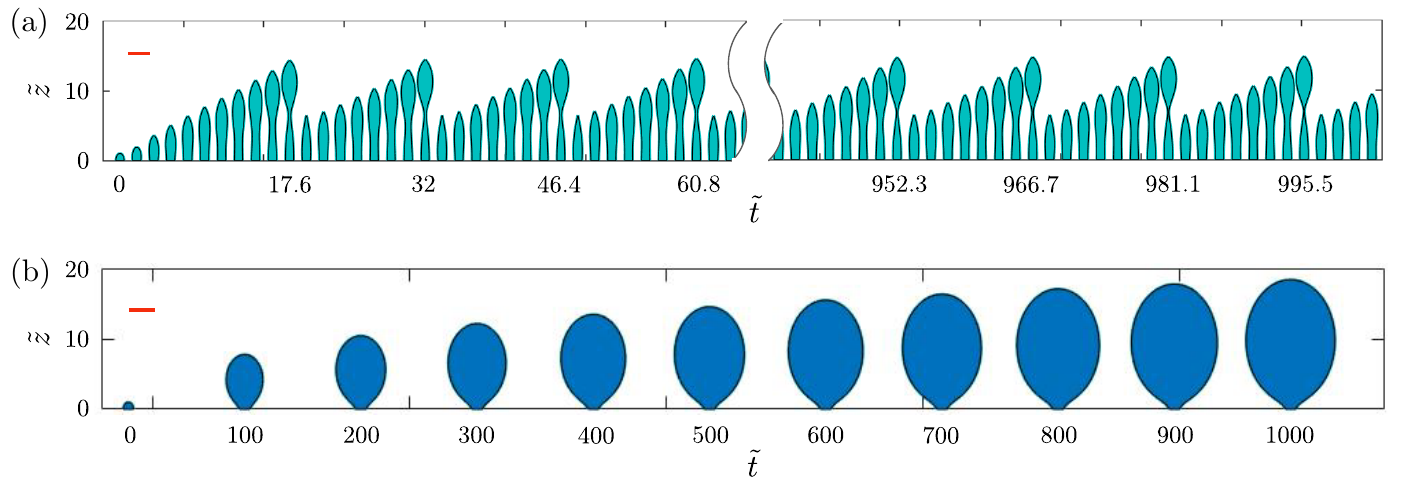}
\caption{Comparison of the nonlinear behaviors of (a) equations \eqref{eqN2} for a silicon-in-silica fiber at $\mathit{We} = 0.01$ and constant $\mathit{Ca} = 1$, and (b) equations \eqref{appEq1} for a jet in an inert medium at $\mathit{We} = 0.01$ and $\mathit{Ca} = 1/3$. The parameter values are chosen such that the two systems share the same dispersion relation, with their only difference being the nonlinear form of the viscous term. In both cases, the dimensionless interface radius $\tilde{h} = h/h_0$ is plotted at fixed time intervals, and the red bar corresponds to a horizontal length scale of 5 dimensionless units.}
\label{HolViscComp}
\end{figure}

The purpose of this appendix is to compare the nonlinear behavior of equations \eqref{eqN2} for a silicon-in-silica fiber at constant capillary number with that of equations \eqref{appEq1} for a jet in an inert medium. We consider the low Weber number limit $\mathit{We} = 0.01$, and use $\mathit{Ca} = 1$ and $1/3$ for \eqref{eqN2} and \eqref{appEq1}, respectively, in such a way that the linear dispersion relations of the two systems are identical; the only difference between them resides in the nonlinear form of the viscous term. In \eqref{eqN2}, the viscous term originates from the radial velocity of the outer silica, while in \eqref{appEq1} it is due to the axial velocity of the jet. In both cases, we start from a hemispherical shape $h/h_0=(1 - \tilde{z}^2)^{1/2}$ and we numerically compute the evolution of the system over a thousand nondimensional time units, using a spatial grid size $d\tilde{z} = 0.04$. Figure \ref{HolViscComp} shows the resulting cascade plots of the dimensionless interface radius $\tilde{h} = h/h_0$ at fixed time intervals for (a) equations \eqref{eqN2} describing the silicon-in-silica fiber and (b) equations \eqref{appEq1} describing the viscous jet. 
Surprisingly, equations \eqref{eqN2} for the silicon-in-silica fiber produce regularly-spaced droplets, while equations \eqref{appEq1} for the viscous jet lead to the formation of one ever-growing pendant drop. The corresponding videos are included in the Supplementary Materials.

\bibliography{Draft}

\end{document}